\documentclass[preprint,12pt]{elsarticle}

\usepackage{amssymb}
\usepackage{amsmath}

\newcommand{\refEq}[1]{Eq.~\ref{#1}}
\newcommand{\refF}[1]{Fig.~\ref{#1}}

\newcommand{\refS}[1]{section~\ref{#1}}
\newcommand{\refT}[1]{table~\ref{#1}}
\newcommand{\refApp}[1]{appendix~\ref{#1}}

\usepackage{hyperref}

\usepackage{color}

\journal{Physics of the Dark Universe}

\begin{document}

\begin{frontmatter}



\title{Analytical $poly\Lambda$CDM dynamics} 


\author[a]{Pierros Ntelis}
\author[b,c]{Jackson Levi Said}

\affiliation[a]{Independent Research Affiliation, Former Aix Marseille Univ., Marseille, France}
\affiliation[b]{Institute of Space Sciences and Astronomy, University of Malta, Msida, Malta}
\affiliation[c]{Department of Physics, University of Malta, Msida, Malta}            

\begin{abstract}
We develop a novel analytical dynamical analysis to derive precise energy density ratio evolutions for the $\phi$CDM and $poly\Lambda$CDM models, comparing them to the standard  $\Lambda$CDM model and validating against numerical solutions. Analytical solutions for the quintessence, i.e. $\phi$CDM, show sub-\% agreement with  $\Lambda$CDM with greater reliability than numerical integration of stiff systems. The $poly\Lambda$CDM model, a phenomenological modified gravity framework, captures radiation, matter, dark energy, and exotic epochs, offering a streamlined yet comprehensive alternative to existing studies. Its dynamics reveal a \textit{global transition from a dark energy-dark matter exchange reflector, through saddle points of matter, radiation, curvature, and modified gravity, to an SVT modified gravity attractor-saddle, and finally to a cosmological constant attractor in the far future, with saddle transitions between modified gravity components}. The $poly\Lambda$CDM model integrates modified gravity models, using dynamical analysis to distinguish observationally viable critical points and differentiate gravity epochs. All three models align with observed cosmic evolution, but $poly\Lambda$CDM’s richer phenomenology provides deeper insights into modified gravity dynamics. Code available at \href{https://github.com/lontelis/Analytical_polyLCDM_dynamics}{GitHub}.
\end{abstract}


\begin{highlights}
\item Research highlight 1: Novel analytical dynamical analysis method development
\item Research highlight 2: Novel phenomenological model that encapsulate several modified gravity models
\item Research highlight 3: Method to disentangle phenomenological aspects of modified gravity
\end{highlights}


\end{frontmatter}



\tableofcontents

\section{Introduction}\label{Introduction}

In modern cosmology, most extensions of gravity are formulated by modifying components of the action 
\cite{CAPOZZIELLO2011167,2018FrASS...5...44E,2016PhR...633....1P,2012PhR...513....1C,PerenonMarinoniPiazza,carrasco2012effective,1998PhLB..436..257A,1999PhRvL..83.3370R,2014IJMPD..2343009B,2000PhLB..485..208D,Acunzo:2021gqc,Capozziello:2021krv,Capozziello:2018gms,Zwane:2017xbg,Das:2023hbw}. Such modifications typically involve alterations to either the geometric sector or the matter Lagrangian, aiming to address cosmological tensions \cite{Schoneberg:2021qvd,2023FoPh...53...29N,Ntelis:2024fzh} and other challenges inherent to the concordance model. In this work, we consider a homogeneous and isotropic universe \cite{Ntelis:2016suu,Laurent,Ntelis:2016utg,2017JCAP...06..019N,Ntelis:2018tlj} to explore a class of models designed to alleviate cosmological tensions \cite{Schoneberg:2021qvd}. As community, we investigate models within Riemannian geometries, a topic of considerable recent interest \cite{CANTATA:2021ktz}. 

The urgent need for a comprehensive understanding of cosmological dynamics necessitates a thorough examination of various cosmological models. We have previously analyzed the dynamics of standard $\Lambda$CDM, $\phi$CDM \cite{Ntelis_Said:2024,Ntelis_Said:2024b}, and quadratic functors of action theories in the context of the $\Lambda$CDM model \cite{Ntelis_Said:2024c}. Building upon these foundations, this study focuses on analytical approaches to elucidate the evolution of cosmological systems, providing crucial insights when compared to numerical methods. 

\textit{The motivation of the study is that we develop a novel analytical dynamical analysis to obtain faster and accurate dynamical analysis energy density ratio evolution solutions and compare them to the numerical ones. Then we apply the analytical dynamical analysis on the known quintessence model, i.e. $\phi$CDM model, and compare the energy density evolutions to the standard numerical dynamical analysis energy density evolutions found in the literature to validate the analysis. Then we develop the novel $poly\Lambda$CDM model, which effectively captures a set of modified gravity models, and we apply the novel analytical dynamical analysis on the model in order to study its dynamics.}

While significant efforts have been devoted to exploring modifications to the geometric aspects of gravitational theories \cite{CANTATA:2021ktz,Bahamonde:2017ize,Bahamonde:2020lsm,Bahamonde:2021akc,Aoki:2023sum,Bahamonde:2021gfp}, this work does not focus on such extensions. Additionally, alternative approaches, including $f(R)$ theories, Horndeski gravity, non-Riemannian cosmologies, and functor of actions theories \cite{2023FoPh...53...29N,Ntelis:2024fzh,sym17050777}, lie beyond the scope of our analysis. Instead, we aim to construct a phenomenological model, in which we add cosmological species, that are design to have some non degenerate scaling dependence of the energy density ratios. Therefore, our motivation lies in post-parametrised Friedman (PPF) equations and post-parametrised Newton (PPN) models \cite{Hu:2007pj,Baker:2012zs}. 

The $\phi$CDM model has been studied extensively in the literature \cite{Copeland:1997et,Singh:2015rqa,vandeBruck:2016jgg,Brahma:2019kch,Hossain:2023lxs,Ramadan:2023ivw, Myrzakulov:2025yxy,Dinda:2025iaq,Lin:2025gne,Li:2025cxn}. Building on previous works \cite{Copeland:1997et,Copeland2006}, who analyze a single barotropic fluid model, our work distinguishes between the energy densities of matter and radiation. Similarly, while \cite{vandeBruck:2016jgg} discuss interacting dark energy models with some focus on $\phi$CDM dynamics, we provide a more detailed and specific exploration while we are building on the $\phi$CDM framework. Furthermore, compared to \cite{Ramadan:2023ivw}, who consider three fluids but prioritize radiation energy density, we focus on the matter energy density and separately account for the other exotic energy densities. 

To that end, we introduce the $poly\Lambda$CDM model which is a phenomenological modified gravity model. Our analysis presents a complete set of critical points, numerical solutions, and detailed insights into the potential configurations within a novel $poly\Lambda$CDM dynamical system.

In this study, we perform two comparative dynamical analysis. One between $\Lambda$CDM and $\phi$CDM models and one between $poly\Lambda$CDM and $\Lambda$CDM models.

In particular, we perform a comparative dynamical analysis of the $\Lambda$CDM and $\phi$CDM models by solving analytically both models, and by constructing three-dimensional dynamical systems with three variables for finding numerical evolutions. This approach enables us to interpret and compare the underlying dynamics more effectively. A key difference from earlier studies \cite{Ntelis_Said:2024} is that we treat the continuity equations for the kinetic and potential terms separately. This simplification eliminates the need for additional variables such as $\lambda$ and $\Gamma$, often used in similar analyses. Note that a potential is not defined and that the treatment is general in that respect. In essence this is a comparative analysis with numerical dynamical analyses found in the literature. Our analytical approach shows that we can have an analytical dynamical analysis validation in cosmological model building.

Furthermore, and more importantly, we perform a comparative dynamical analysis of the  $poly\Lambda$CDM and $\Lambda$CDM models by solving analytically both models, and by constructing seven-dimensional dynamical systems with seven variables for finding numerical evolutions. This approach enables us to interpret and compare the underlying dynamics more effectively. 

Our investigation focuses on the physics governing the universe's background evolution during both early and late epochs. At early times, we analyze initial conditions and inflationary dynamics, while at late times, we examine the accelerated expansion and large-scale structure formation. To address existing cosmological tensions, we simulate equations governing motion, energy densities, and characteristic scales of the universe. These efforts aim to lay a foundation for exploring non-Riemannian cosmologies. Key observables include the ratios of matter, radiation, and dark energy densities, as well as the Hubble expansion rate. By reanalyzing data from surveys such as Euclid \cite{EUCLID:2020jwq,Euclid:2021qvm,Euclid:2024yrr} and DESI \cite{Alam:2020jdv}, we aim to test the robustness of these models.


We also note that poly$\Lambda$CDM appears to be the first model to extend $\Lambda$CDM with seven components---matter ($\rho_m \propto a^{-3}$), radiation ($\rho_r \propto a^{-4}$), curvature ($\rho_k \propto a^{-2}$), cosmological constant ($\rho_\Lambda \propto a^{0}$), and three additional terms $x$ ($\rho_x \propto a^{-1}$), $z$ ($\rho_z \propto a^{-5}$), and $v$ ($\rho_v \propto a^{-\alpha}$)—while providing analytical solutions for their dynamics (Sections 2.3.1, A.1). Other studies, such as those on quintessence or multi-fluid systems, typically use fewer components or rely on numerical methods without these exact scalings (e.g., \cite{Copeland:1997et,Bahamonde:2017ize,Azreg-Ainou:2013jxa}). This setup, though basic, offers a new way to look at cosmic evolution with a full set of solutions, which we included as a starting point.


Ultimately, this work seeks to provide a more comprehensive cosmological model by integrating Riemannian geometry, exotic dark energy, and gravity modification components. By addressing current tensions and discrepancies, we aim to contribute to a deeper understanding of the universe's fundamental properties. Motivated by ongoing tensions such as the Hubble tension and $\sigma_8$  discrepancy, we introduce $poly\Lambda$CDM to explore deviations from $\Lambda$CDM that may reconcile these issues.

This paper is structured as follows: 
In \refS{sec:Dynamical_Analysis_phiLambdaCDM}, we describe the dynamical analysis of the $\phi$CDM model.
In \refS{sec:Time_epochs_phiCDM}, we present the distinct epochs of $\phi$CDM model, and we identify all the energy density ratio equalities. 
In \refS{sec:Comparison_phiLCDM}, we present the analytical solutions of $\phi$CDM, and we compare them to numerical ones, and analytical solutions of $\Lambda$CDM.
In \refS{sec:polyLCDM_analysis}, we describe the analysis of the $poly\Lambda$CDM cosmology.
In \refS{sec:from_action_to_friedman_equations}, we describe how we obtain the Friedman equations from the action of the model.
In \refS{sec:physical_interpretation}, we describe the possible interpretation of the model.
In \refS{sec:dynamical_analysis}, we describe the dynamical analysis applied to the system, and we provide our results.
Finally, in \refS{sec:conclusion_discussion}, we conclude and discuss our results.

\section{Analysis of $\phi$CDM cosmology}\label{sec:Dynamical_Analysis_phiLambdaCDM}

In this section, we describe the dynamical system of $\phi$CDM, and then we describe the analytical solutions found. The $\phi$CDM model is built on an action which has minimally coupled dynamical scalar potential, $\phi$, which is a scalar dynamical field, with a kinetic term, described by partial derivatives, $\partial_t \phi$, and $V=V[t, \Lambda; \phi(t)]$ is its potential. Then by applying the least action principle, $\delta S_{\phi\text{CDM}}=0$, we get the following set of equations.
The $\phi$CDM Friedmann equations are 
\begin{eqnarray}
	\label{eq:Friedman_1st_compact_phiLambdaCDM}
	3H^2(t) &=& \kappa^2 \sum_{s \in \left\{ m, r, \phi \right\}} \bar{\rho}_{\rm s}(t) \\
	\label{eq:Friedman_2nd_compact_phiLambdaCDM}	
	2\dot{H}(t) + 3H^2(t) &=& - \kappa^2\sum_{s \in \left\{ m, r, \phi \right\}} w_s(t) \bar{\rho}_{\rm s}(t) 
\end{eqnarray}
where $H(t)$ is the hubble expansion rate, $s$ is index of the species, in our case, $m$ matter, $r$ radiation, $\phi$ dynamical scalar field, where $\kappa^2 = 8\pi G_{\rm N}/c^4$, where $G_{\rm N}$ is the Newton gravitational constant, and $c$ the speed of light.
The Klein-Gordon equation for the dynamical scalar field, $\phi$, is written as
\begin{eqnarray}
	\ddot{\phi} + 3H \dot{\phi} +  V_{, \phi} = 0
\end{eqnarray}
where $V_{, \phi}=\partial_\phi V(\phi)=\frac{dV(\phi)}{d\phi}$ is the derivative of the potential, in respect of the dynamical scalar field.

\subsection{Analytical dynamical solutions}\label{sec:dimensionless_variables_and_representative_3DphiLambdaCDM}

We use the equations of states are well known for these fields, and fixed to the following values, $\left\{ w_m, w_r \right\} = 	\left\{ 0, 1/3\right\}$, while the scalar-field equation of state, $w_\phi = w_{\phi} (x,v)= \frac{x - v}{x + v}$.

We define the following dimensionless variables
\begin{align}
 m= \Omega_{\rm m} &= \frac{\kappa^2 \bar{\rho}_m(t)}{3 H^2(t)} \quad , \quad r = \Omega_{\rm r} =  \frac{\kappa^2 \bar{\rho}_r(t)}{3 H^2(t)} \\
x=  \Omega_{ x}  
&= \frac{\kappa^2}{3 H^2}X =  \frac{\kappa^2}{3 H^2} \frac{\dot{\phi}^2}{2} =  \frac{\kappa^2 \dot{\phi}^2 }{6 H^2}   \\ 
v=  \Omega_{ v}
&= \frac{\kappa^2}{3 H^2}V[t;\phi(t),\Lambda] \\
 \tilde{\phi} = \Omega_{\tilde{\phi}} &=  \Omega_{ x} +  \Omega_{ v}
\end{align}

Note that the effective equation of state is written as 
\begin{eqnarray}
	w_{\rm eff} 
	&= \sum_{s \in \left\{ m,r, x,y\right\}} w_s(t) \Omega_s(t) = 
	\frac{1}{3} r(t) + x(t) - y(t)
\end{eqnarray}

With a specific choice of the parameters of the modelled guided by the choice of the potential, we end up with a representative 3D $\phi$CDM model, since we use the first Friedman equation to define the radiation evergy density ration as a function of the rest of the energy density ratios.

Note that by considering separately the continuity equation for the kinetic term, and the potential term, we end up to a new system, which is simpler than previous studies.

We have the equations: 
\begin{align}
	 1 &= m + r + x + v \\
           f &= 3 + r + 3 x  - 3 v 	 \\	
	m' &=  m (f-3 )   \\	
	 r' &= r  (f-4)   \\
	x' &= x (f-3 )   \ \\	
	v' &=  v  f 
\end{align}


The set of equations have exact solutions which can be written as
\begin{align}
	\Omega_m  &= \Omega_{m0} e^{-3N}  \left[ e^{-3N} \Omega_{m0}  + e^{-4N} \Omega_{r0} + e^{-3N} \Omega_{x0}  +  \Omega_{v0} \right]^{-1}   \\	
	\Omega_r  &= \Omega_{r0} e^{-4N}  \left[ e^{-3N} \Omega_{m0}  + e^{-4N}  \Omega_{r0} + e^{-3N} \Omega_{x0}  +  \Omega_{v0}\right]^{-1}   \\	
	\Omega_x  &=\Omega_{x0} e^{-3N}  \left[ e^{-3N} \Omega_{m0}  + e^{-4N}  \Omega_{r0} + e^{-3N} \Omega_{x0}  +  \Omega_{v0} \right]^{-1} \\
	\Omega_v &= \Omega_{v0} \left[ e^{-3N} \Omega_{m0}  + e^{-4N}  \Omega_{r0} + e^{-3N} \Omega_{x0}   +  \Omega_{v0} \right]^{-1} 
\end{align}

To get the solutions, we separate the variables, and then work the equations two by two, by eliminating the term f, which creates solvable differential equations.
Then we combine all solutions to the functional form of $f$, and we substitute that to one of these equation, to find the solvable form for one of the solutions, i.e. the $r$ variable function. Then the rest of the functions depend on that one solution. These actions makes the system solvable analytically. You can find the exact calculations in the supplementary material.

These solutions have the following limiting behaviours.
The radiation domination epoch in the far past, is
\begin{align}
	\Omega_r( N \to -\infty  ) &= 1  \\
	 \Omega_x( N \to -\infty  ) &=   0 = 	
	\Omega_v( N \to -\infty  ) 	 =
	\Omega_m( N \to -\infty  )  
\end{align}
today we have
\begin{align}
	\Omega_r( N \to 0  ) &= \Omega_{r,0}  \\
	 \Omega_x( N \to 0  ) &=   \Omega_{x,0} \\	
	\Omega_v( N \to 0  ) &=   \Omega_{v,0} \\	 
	\Omega_m( N \to 0  )  &=    \Omega_{m,0}	  
\end{align}
and the cosmological constant domination epoch in the far future, is
\begin{align}
	\Omega_m( N \to +\infty  ) &= 0 = 
	\Omega_r( N \to +\infty  )  =   
	 \Omega_x( N \to +\infty  ) \\		
	\Omega_v( N \to +\infty  ) &=   1 	
\end{align}

Note that this system becomes the standard $\Lambda$CDM, when $\Omega_x(N) \to 0$ and $\Omega_v(N) \to \Omega_\Lambda(N)$. We describe the model in the \refApp{sec:LCDMcosmology_analytics}

\subsection{Time epochs in $\phi$CDM model}\label{sec:Time_epochs_phiCDM}

Given the aforementioned evolution of the energy densities ratios of the $\phi$CDM model, we identify the following important epochs. We illustrate them in \refT{tab:epoch_time_triplet_different_energy_density_ratio_species_equivalences_different_values}. As observed in here, we have identified different epochs, depending on the equality of energy densities. We have identified the usual epochs, such as today, initial, matter-radiation equality, recombination, decoupling, and we have discovered further the epochs rising from the dynamical dark energy (DDE), i.e. kinetic-DDE, potential-DDE, kinetic-potential, radiation-DDE, matter-DDE epochs.

\begin{table}[h!]
\centering
\begin{tabular}{|l|c|c|c|}
\hline
\textbf{Epoch} & lapse function, $N(t)$ & redshift, $z(t)$ & scale factor, $a(t)$ \\ \hline
today, $0$ & 0 & 0 & 1 \\ \hline
matter-DDE, $m \tilde{\phi}$ & -0.251 & 0.286 & 0.778 \\ \hline
radiation-DDE, $r \tilde{\phi}$ & -2.04 & 6.68  & 0.130 \\ \hline
Kinetic-Potential, $xv$ & -3.292 & 25.9 & 0.0372 \\ \hline
decoupling & -7.01 & 1089 & $9 \times 10^{-4}$ \\ \hline
recombination & -7.01 & 1100 & $9 \times 10^{-4}$ \\ \hline
matter-radiation, $m r$ & -8 & 2999 & $3 \times 10^{-4}$ \\ \hline
initial, $i$ & -12 & $2 \times 10^5$ & $6 \times 10^{-6}$ \\ \hline
Kinetic-DDE, $x \tilde{\phi}$  & $-\infty$ & $+\infty$ & $0$ \\ \hline
Potential-DDE, $v \tilde{\phi}$  & $-\infty$ & $+\infty$ & $0$ \\ \hline
\end{tabular}
\caption{\label{tab:epoch_time_triplet_different_energy_density_ratio_species_equivalences_different_values} Classification of time triplet epoch according to species energy density ratio equivalences, for $\phi$CDM. We assume the choice $\Omega_{m,0}=0.32$, $\Omega_{x,0} = 3.5 \times 10^{-5}$ and $\Omega_{v,0} = 0.68$, $\Omega_{r,0}= 2 \times 10^{-4}$. }
\end{table}

\subsection{Comparison between analytical and numerical solutions of $\phi$CDM}\label{sec:Comparison_phiLCDM}

The previous system of equations is simplified to the representative 3D system,

\begin{eqnarray}
	m' &=&  m (1 - m + 2 x  - 4 v)     \\	
	x' &=& x (1 - m + 2 x  - 4 v)\\	
	v' &=&  v \; (4 - m + 2 x  - 4 v)
\end{eqnarray}
where the \( r=1-m-x-v\) . 

\begin{figure*}[h!]
    \hspace{-2cm}
    \centering 
    \includegraphics[width=160mm]{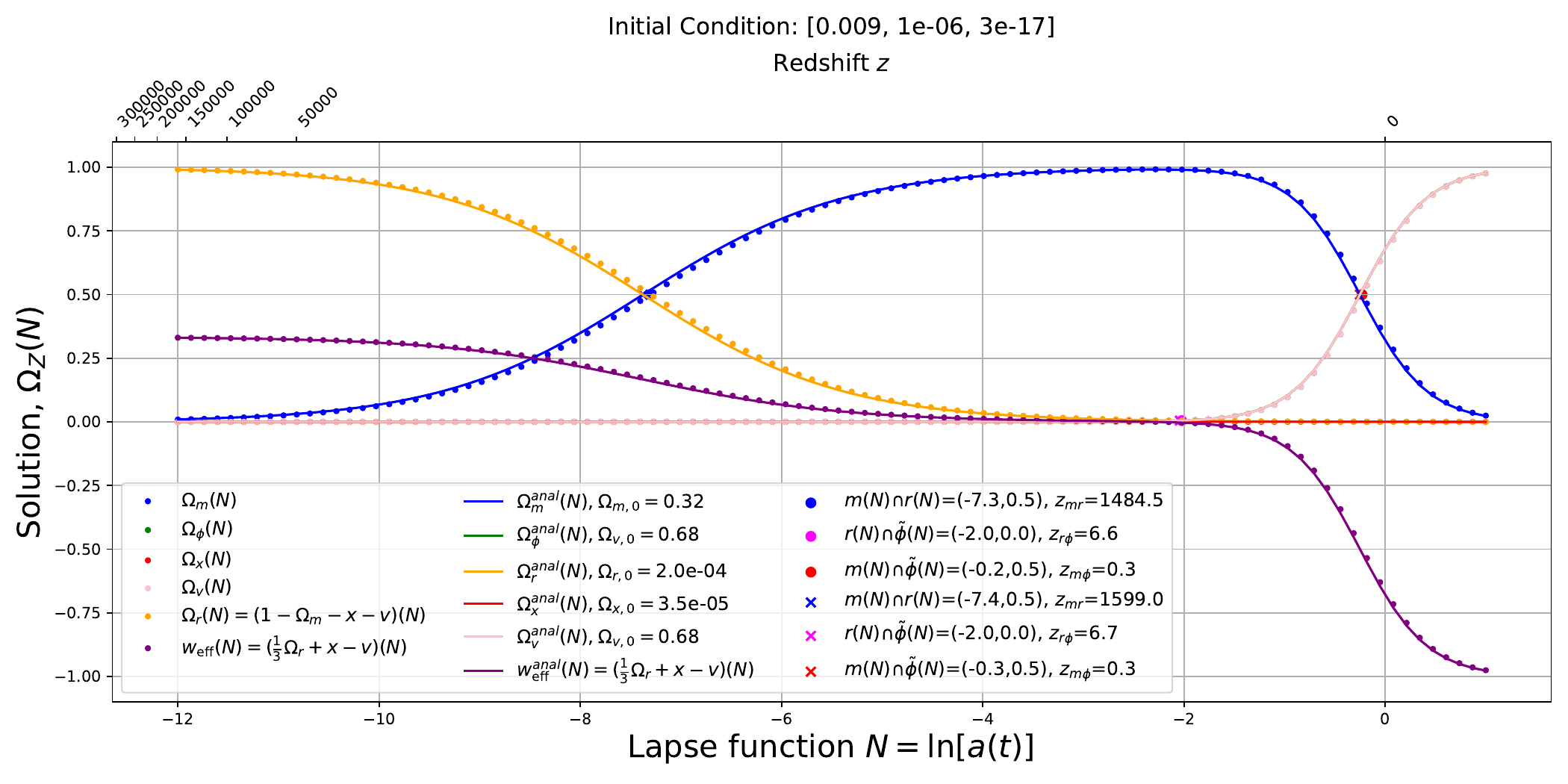}            
    \caption{\label{fig:ChatGPT_phiLCDM_7D_3D_set_of_sim_diff_eqns_num_anal_solutions}   We illustrate the analytical with a constant line and the numerical solutions with dotted line of the model of $\phi$CDM cosmology, in the case which $\texttt{var-zero-is}=3 \times 10^{-17}$. 
    In the third column in the label we describe the intersection point between two function variables, and the corresponding redshift, for example
$m(N) \cap r(N) = (-7.3, 0.5)$, $z_{\rm mr} = 1484.5$ means that matter and radiation meet at $N=-7.3$, in an energy density of matter-radiation equality $\Omega_{\rm mr} = 0.5$, at a corresponding redshift of matter-radiation equality $z_{\rm mr} = 1484.5$  
    [See \refS{sec:Comparison_phiLCDM}] }
    \end{figure*}

We solve numerically\footnote{We use the \texttt{scipy.integrate.solveivp} to solve the system.} the 3D system of differential equations.
We find the following numerical solutions and we present them in the \refF{fig:ChatGPT_phiLCDM_7D_3D_set_of_sim_diff_eqns_num_anal_solutions}, for several specified conditions. We find the $\phi$CDM results are similar to the $\Lambda$CDM model, on the behaviour of the epoch evolution of the different species of the model. What changes is the gradient  behaviour of dark energy as expected.

In particular, we find that the radiation is dominant in the past, and decays as time passes. The matter starts significantly low, has a peak, during intermediate times, and then decays towards the today's epoch, and the far future. Note that the pink curve is superimposed to the green curve, which show that the potential term matches quite close with the one of the total behaviour of the dynamical dark energy density ratio.

In the third column, in the label of \refF{fig:ChatGPT_phiLCDM_7D_3D_set_of_sim_diff_eqns_num_anal_solutions}, we describe the intersection point between two function variables, and the corresponding redshift, for example
$m(N) \cap r(N) = (-7.3, 0.5)$, $z_{\rm mr} = 1484.5$ means that matter and radiation meet at $N=-7.3$, in an energy density of matter-radiation equality $\Omega_{\rm mr} = 0.5$, at a corresponding redshift of matter-radiation equality $z_{\rm mr} = 1484.5$. In general this is described as 
\begin{align}
 	X(N) \cap Y(N) = (N_0, \Omega_0) \Leftrightarrow z_{\rm XY} = z_0 \; 	, X,Y \in \left\{ m, r, \tilde{\phi} \right\} \; , \; N_0, \Omega_0, z_0 \in \mathbb{R}^3
\end{align}
where $X,Y \in \left\{ m, r, \tilde{\phi} \right\} \; , \text{and} \; N_0, \Omega_0, z_0 \in \mathbb{R}^3$.
This mathematical phrase translates to
that the intersection point between variable function $X$, and variable $Y$, has coordinates in the lapse function, and energy density plane $(N_0, \Omega_0)$, and corresponds to the redshift of the equality epoch between the epoch of $X$, and epoch of $Y$, while the epochs that we have in this model are matter, $m$, radiation, $r$, and dynamical field, $\tilde{\phi}$. Therefore in \refF{fig:ChatGPT_phiLCDM_7D_3D_set_of_sim_diff_eqns_num_anal_solutions}, we present all such points of epoch equalities.

\begin{figure*}[h!]
    \centering 
    \includegraphics[width=160mm]{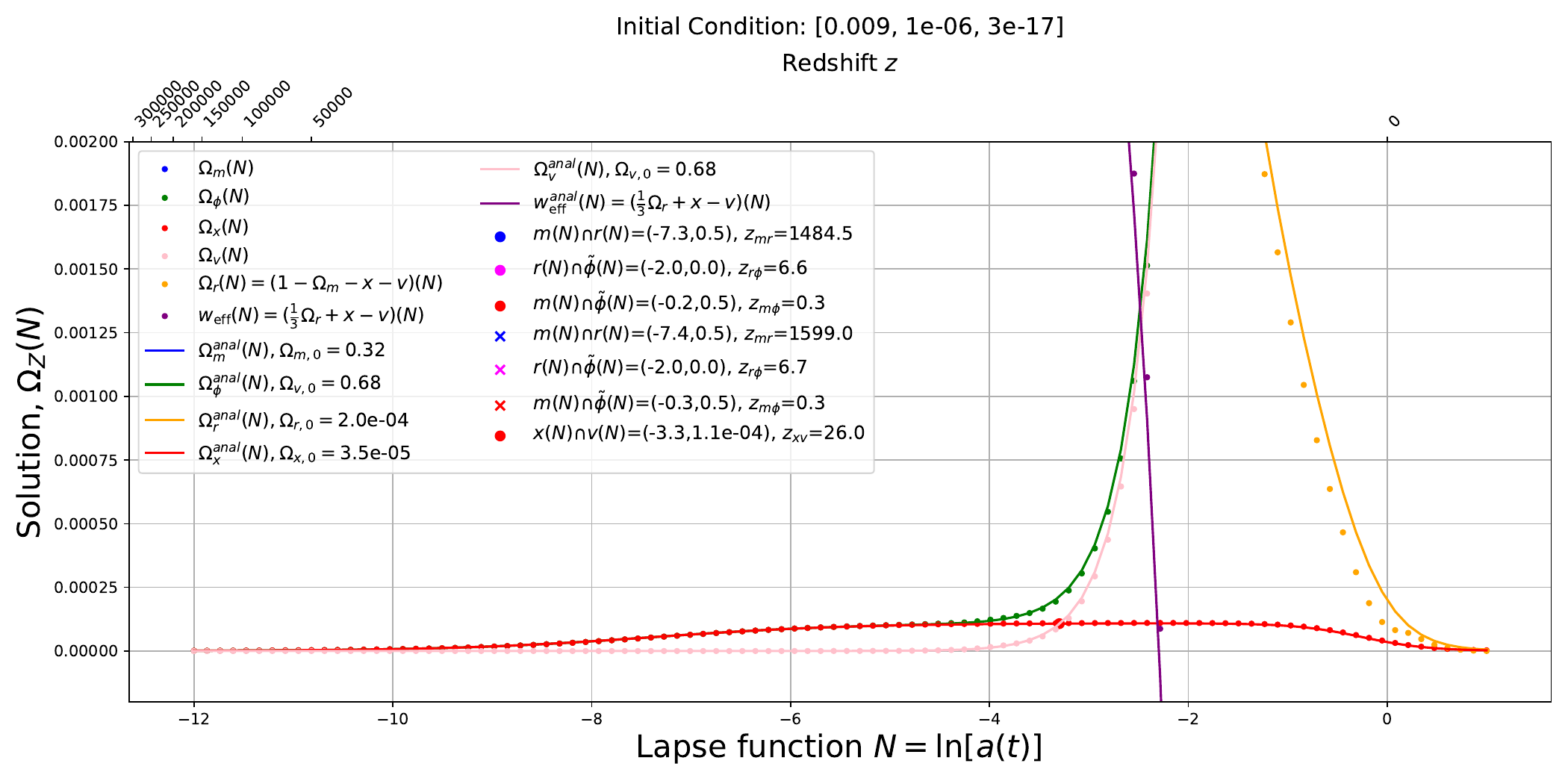}    
        \caption{\label{fig:ChatGPT_phiLCDM_7D_3D_set_of_sim_diff_eqns_num_anal_solutions_ZOOMIN} As \refF{fig:ChatGPT_phiLCDM_7D_3D_set_of_sim_diff_eqns_num_anal_solutions},  but we zoom in so that we observe clearly the evolution of the scalar kinetetic term. We illustrate the analytical with a constant line and the numerical solutions with dotted line of the model of $\phi$CDM cosmology, in the case which $\texttt{var-zero-is}=3 \times 10^{-17}$. [See \refS{sec:Comparison_phiLCDM}] }
    \end{figure*}

We observe the subtle effects of kinetic and potential terms in the \refF{fig:ChatGPT_phiLCDM_7D_3D_set_of_sim_diff_eqns_num_anal_solutions_ZOOMIN}, where we zoom in the region. We have the following color-coding for the energy density ratios.
Blue is matter, green is dynamical dark energy, red is kinetic term, pink is potential term, yellow is radiation, and magenta is the effective equation of state. We also mark the equality epoch coordinates, as well as their corresponding redshifts.

We find that the potential kinetic term starts significantly low, but it leads to a dominant scalar kinetic potential epoch in the far future, which signifies a de Sitter universe, i.e. a universe filled with the cosmological constant.  We observe a mild increase of the scalar kinetic term energy density, in the late universe, i.e. about  $N_{x} \in [ -6, -1 ]$, i.e. $z_{x} \in [ 402, 1.71 ]$, and then a decay to 0, towards the future. Note that the pink curve is superimposed to the green curve, in most regions, which show that the potential term matches quite close with the one of the total behaviour of the dynamical dark energy density ratio. However, near the peak of the kinetic term there is a difference, and we can see the evolution of all three ingredients, kinetic, potential and dynamical dark energy density.

\begin{figure*}[h!]
    \centering 
    \includegraphics[width=160mm]{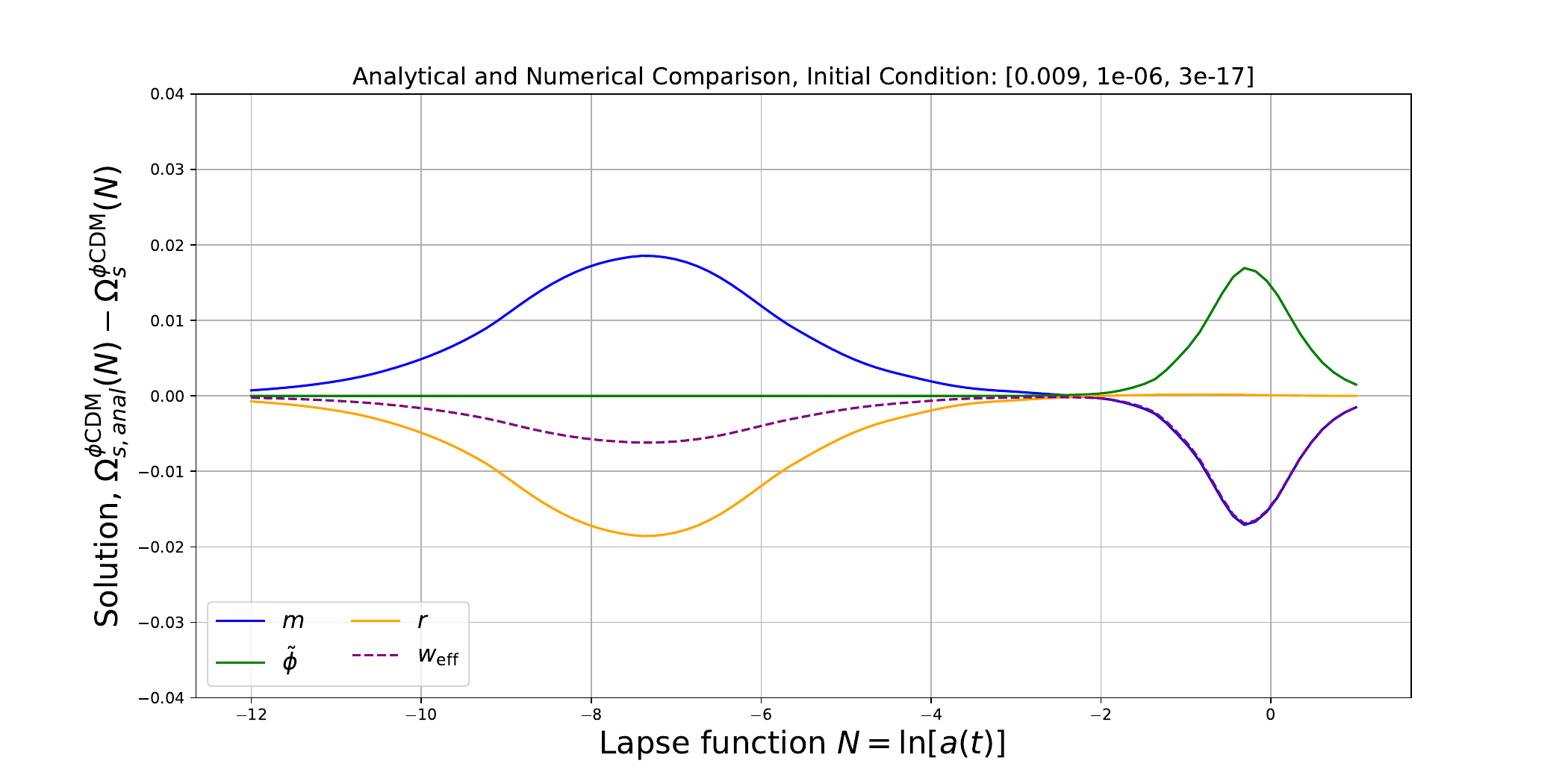}        
    \caption{\label{fig:ChatGPT_phiLCDM_comparison_analytical_numerical}  We present the comparison between the analytical and the numerical evolution of the different species of energy density ratios as a function of the lapse function, $\Omega_s(N)$.  [See \refS{sec:Comparison_phiLCDM}] }
    \end{figure*}

In \refF{fig:ChatGPT_phiLCDM_comparison_analytical_numerical}, we present the comparison between the analytical and the numerical evolution of the different species of energy density ratios as a function of the lapse function, $\Omega_s(N)$. We find a good agreement between the analytical and numerical solutions, which lies in less than 10\% differences. The differences are due to numerical choices of the initial and and today's conditions for the energy density, and also the correspondence of negative infinity with the value $N=-12$.

\begin{figure*}[h!]
    \centering 
    \includegraphics[width=160mm]{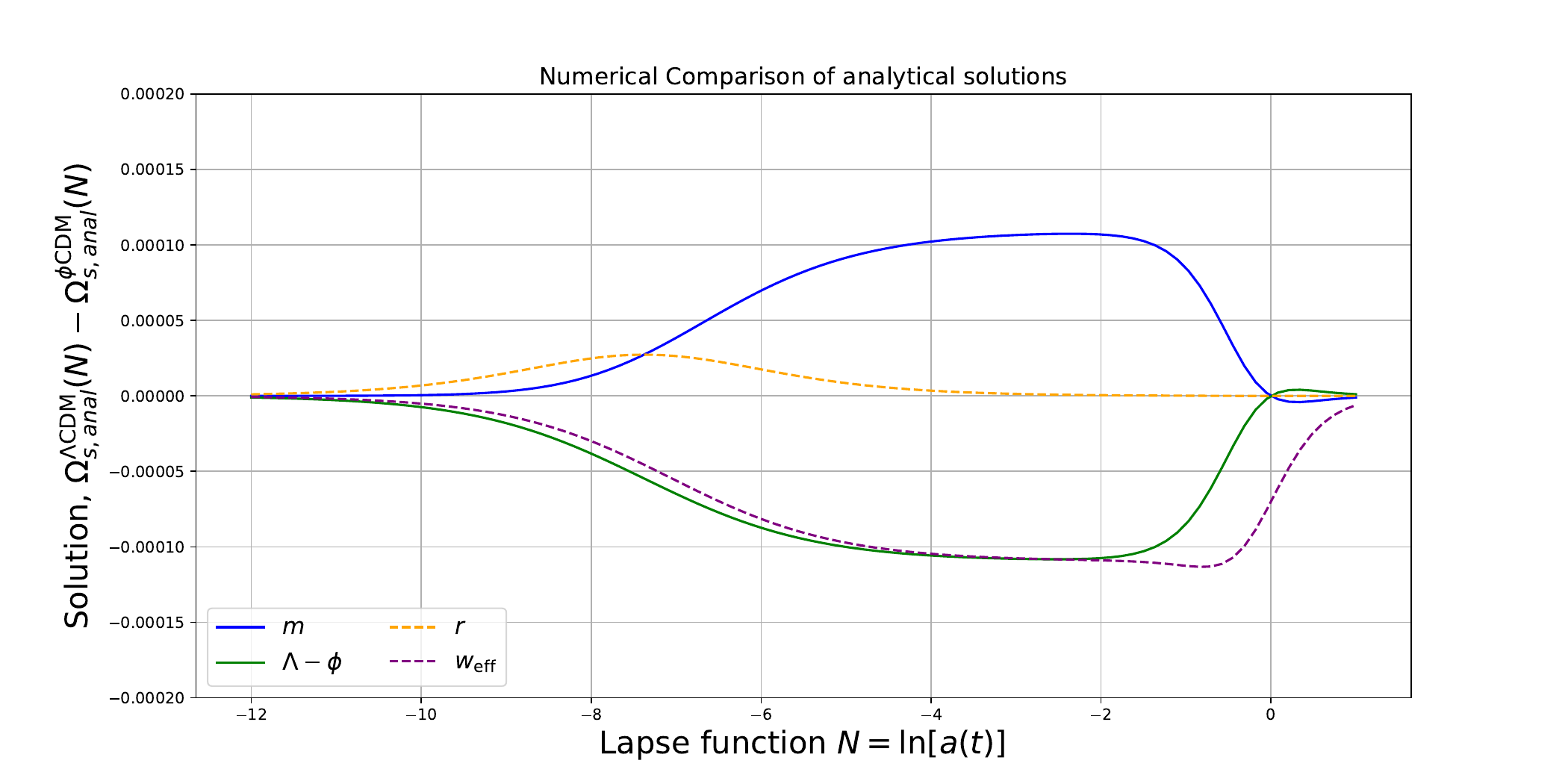}   
     \caption{\label{fig:ChatGPT_phiLCDM_vs_LCDM_comparison_analytical_ZOOMIN} Zoomed in analytical comparison of energy density evolution of $\phi$CDM vs $\Lambda$CDM. [See \refS{sec:Comparison_phiLCDM}] }
    \end{figure*}

In \refF{fig:ChatGPT_phiLCDM_vs_LCDM_comparison_analytical_ZOOMIN},  we zoom in and we present the numerical comparison between the analytical evolution of the $\Lambda$CDM model and and $\phi$CDM models, and we have zoomed in the sub-\% level. We find that most species agree at sub-\% level. However, as expected there is a peak of the energy density ratio evolution for matter and dark energy densities. In particular there is less matter energy density ratio  and and more dark energy density ratio  in the $\phi$CDM model, rather than the ones in the $\Lambda$CDM model.

\subsection{Overall comparison of analytical and numerical energy densities}
The sub-\% differences between the analytical energy density evolutions of $\Lambda$CDM and  $\phi$CDM indicate close agreement between the models. In contrast, the larger differences between the analytical and numerical solutions for  $\phi$CDM, despite solving the same equations, suggest that the analytical evolutions are more precise and less susceptible to computational errors than the numerical ones in this implementation. This highlights the challenges of numerically integrating stiff systems and supports the use of analytical solutions for reliable results when available.

\subsection{Note on the components of $\phi$CDM and $poly\Lambda$CDM}

Note that the components $\Omega_x$ and $\Omega_v$ of $\phi$CDM are different than the components $\Omega_x$ and $\Omega_v$ of $poly\Lambda$CDM, which are analysed to the next section.

\section{Analysis of $poly\Lambda$CDM cosmology}\label{sec:polyLCDM_analysis}

In this section, we describe the analysis applied on the $poly\Lambda$CDM cosmology. 

\subsection{From action to Friedman equations}\label{sec:from_action_to_friedman_equations}

The Friedmann equations describe the dynamics of the scale factor \(a(t)\) in a universe described by the Friedmann-Lemaître-Robertson-Walker (FLRW) metric, with curvature. Here, we include the spatial curvature explicitly and derive both the first and second Friedmann equations from first principles.

The total action for this model can still be written in a general form is
\begin{eqnarray}
S = c^3  \int d^4x \sqrt{-g} \, \left[ \frac{R - 2 \Lambda}{2\kappa^2} + \mathcal{L}_{fluid} \right],
\end{eqnarray}
where $c$ is the speed of light, $\kappa^2 = 8\pi G_{\rm N}/c^2$, where $G_{\rm N}$ is the Newton gravitational constant, $g$ is the determinant of a generic metric, and \(\mathcal{L}_{fluid} \) represents the Lagrangian contributions from the different curvature corrections or matter-energy species components. We provide the derivation of all Lagrangian components in \refApp{sec:Lagrangians_new}.

The FLRW metric for a homogeneous and isotropic universe is:
\begin{eqnarray}
ds^2 = -c^2 dt^2 + a^2(t) \left[ \frac{dr^2}{1 - k r^2} + r^2 \left(d\theta^2 + \sin^2\theta \, d\phi^2 \right) \right].
\end{eqnarray}
where  \( k \) is the spatial curvature constant (\(k = -1, 0, +1\) for open, flat, and closed universes, respectively), \( t \) is the cosmic time, and \( r, \theta, \phi \) are the spatial coordinates. The curvature term \( k r^2 \) modifies the spatial geometry. We also define $a(t)$ as the scale factor, $H(t)=\frac{\dot{a}}{a}$ as the hubble expansion rate, where $\dot{a}=\frac{da}{dt}$ . Then we apply the least action principle according to the FLRW accounting for curvature, \( \frac{\delta S}{ \delta g^{\mu\nu}} =0 \).

The $poly\Lambda$CDM Friedmann equations are 
\begin{eqnarray}
	\label{eq:Friedman_1st_compact_polyLambdaCDM}
	3H^2(t) &=& \kappa^2 \sum_{s } \bar{\rho}_{\rm s}(t) \\
	\label{eq:Friedman_2nd_compact_polyLambdaCDM}	
	2\dot{H}(t) + 3H^2(t) &=& - \kappa^2\sum_{s } w_s(t) \bar{\rho}_{\rm s}(t) 
\end{eqnarray}
where each \( \bar{\rho}_s \) evolves according to its equation of state \( w_s \), through \(
\bar{P}_s = w_s \bar{\rho}_s \) .

The dimensionless density parameters for each species, $s$, are defined as:
\begin{eqnarray}
s(t) = \Omega_s(t) = \frac{\kappa^2 \bar{\rho}_s(t)}{3H^2(t)},
\end{eqnarray}

We rewrite the equations as follows.
The total energy density is the sum of contributions from all components:
\begin{eqnarray}
1 = \sum_{s}  \Omega_s(t) = \sum_{s} s(t) .
\end{eqnarray}

The effective pressure determines the acceleration:
\begin{eqnarray}
	f(t)=  -2\dot{H}/H^2 = 3 + 3 \sum_{s } w_s(t) s(t) 
\end{eqnarray}

We consider the generic continuity equation, written in the following equivalent forms for each component, $s$, and we write
\begin{eqnarray}	
	\nabla^{\mu} T_{\mu\nu} &=& 0 \\
	\dot{\rho}_s &=& - 3 (1+w_s) H(t) \rho_s \\
	\Omega_s'(t)  &=& \Omega_s(t) \left[   (- 2) \frac{\dot{H}(t)}{H^{2}(t)} + (-3)(1+w_s) \right] \\
	s'(t)  &=&  s(t) \left[ f - \alpha_s  \right] 		\label{eq:continuity_equations_of_species}			
\end{eqnarray}
where $-\alpha_s = (-3)(1+w_s)$ and $'=\frac{d}{dN}$, where $N = \ln a(t)$ is the lapse time function.

\subsubsection{Friedman equations, generic}
Solving the continuity equations, and integrating once the 2nd Friedmann equation, we get a generic relation between the time and the scale factor, written as  
\begin{equation}
	\boxed{
	t - t_0 =  H_0^{-1} \int_{a_0}^a \frac{da}{a  \left[ \sum_{s } \Omega_{s,0} a^{-\alpha_s  } \right]^{ \frac{1}{2} } }    
	}
\end{equation}
rearranging it, we get the 1st Friedman equation
\begin{equation}
	\boxed{
	H(t) = \frac{\dot{a}}{a} =  H_0  \left[ \sum_{s } \Omega_{s,0} a^{-\alpha_s  } \right]^{ \frac{1}{2} }     
	}
\end{equation}

\subsubsection{Remark on curvature and energy density of curvature}
It is important to remark that we can get an energy density component in the Friedman equation with the scaling of the curvature component, either by introducing curvature to the metric, and varying the action,
\begin{eqnarray}
	g_{\mu\nu} = g_{\mu\nu}(k) \Rightarrow \frac{\delta S}{\delta g_{\mu\nu}(k)}  = 0 \Rightarrow \Omega_k \propto a^{-2} \; ,
\end{eqnarray}
or by having a flat metric, and introducing a term in the action, that has a specific configuration of equation of state that results to a energy density scaling match like the curvature component,

\begin{eqnarray}
	g_{\mu\nu} \neq g_{\mu\nu}(k) \Rightarrow \frac{\delta S( -\rho_k) }{\delta g_{\mu\nu}}  = 0 \Rightarrow \Omega_k \propto a^{-2} \; .
\end{eqnarray}

\subsection{Physical interpretation}\label{sec:physical_interpretation}

The new fields introduced in the model - \(x\),  \(z\), and \(v\) - are associated with new forms of energy densities or components in the universe. We also consider the standard components of matter, $m$, radiation, $r$, curvature, $k$, and cosmological constant, $\Lambda$. Their interpretation depends on the physical context and the role they play in the cosmological dynamics. We provide below possible interpretations and insights into their physical significance.

\subsubsection{Generalized Energy Components}

Each field corresponds to a new type of energy density that contributes to the total energy density of the universe. 
We use the following species
\begin{eqnarray}
	s \;   \in \{ m, r, x, k, z, v, \Lambda \}
\end{eqnarray}
We assign these components to the following initial representations:

- \( m \): matter component of the universe,

- \(  r \): radiation component of the universe,

- \(x\): a form of energy with an unusual equation of state, possibly related to modified gravity or exotic matter.

- \(k\): Curvature of spatial space or an exotic energy density species.

- \(z\): Interacting or coupled dark energy models, where dark matter and dark energy exchange energy or momentum.

- \(v\): A scalar, vector, tensor (SVT) field with specific coupling to gravity, affecting the effective Friedmann equations.

- \(\Lambda\): The cosmological constant, as in standard \(\Lambda\)-dominated cosmology.

Note also that in theories of modified gravity, additional fields often emerge naturally as effective degrees of freedom:

- \(x\), \(k\), \(z\), \(v\) could arise from higher-order corrections to the Einstein-Hilbert action, such as \(f(R)\) gravity, scalar-tensor theories, or extra-dimensional models.

They may also represent effective contributions from the scalar curvature, torsion, or other geometrical properties:

- \(x\): Could be tied to a scalar field coupling to curvature.

- \(z\): Might encode effects from a non-minimally coupled vector field.

- \(v\): Could arise from higher-dimensional terms projected into 4D spacetime.

\subsubsection{Phenomenological Constructs and Observational Predictions}
The fields could serve as phenomenological terms designed to explore deviations from standard cosmology:

- They might encode deviations from the \(\Lambda\)CDM model to fit observational data (e.g., anomalies in CMB or large-scale structure).

- Their parameters (\(\alpha\), \(\beta\), etc.) can be constrained by observations to identify potential new physics.

To interpret these fields concretely, one must examine their observational consequences:

1. Cosmic Microwave Background (CMB): How do \(x\), \(k\), \(z\), and \(v\) affect the CMB anisotropies?

2. Large-Scale Structure (LSS): Do they alter the growth rate of structures or clustering of galaxies?

3. Hubble Parameter Evolution: Can they explain the Hubble tension or deviations in the \(H(z)\) measurements?

4. Gravitational Waves: Do they predict any new modes or effects on gravitational wave propagation?

We leave the answers to these question for future studies, and we focus in the the time scale factor relation and the dynamical analysis of the system.

\subsubsection{Summary}

The fields \(x\), \(k\), \(z\), and \(v\) can be interpreted as contributions from new physics, generalized energy densities, or effective terms arising in modified gravity theories. To fully understand their nature, one needs to:

1. Specify their Lagrangians or equations of state.

2. Investigate their couplings to the metric and other fields.

3. Test their predictions against cosmological observations.

In our work, we do not identify specifically their connection with the lagrangian, nor we test their predictions with current cosmological observations. So we assign to each veriable the initial representation, in order to have a physical interpretation, and we concentrate to the dynamical analysis of the $poly\Lambda$CDM cosmology.
Unlike simpler models, poly$\Lambda$CDM’s seven-component system enables a novel analytical treatment of cosmic dynamics.

\subsection{Dynamical analysis}\label{sec:dynamical_analysis}

In this section, we describe the $poly\Lambda$CDM, and then we describe the dynamical analysis (DA) applied to it. Then by applying the least action principle, $\delta S_{poly\Lambda\text{CDM}}=0$, we get the following set of equations.

Given the energy density ratios for each species, 
we simply write the Friedman equations as
\begin{eqnarray}
	1 &=& \sum_{s \in \left\{ m, r, x, k, z, v, \Lambda \right\}} s (t) \\
	f(t) &=& 3 + 3 \sum_{s \in \left\{ m, r, x, k, z, v, \Lambda \right\}} w_s(t) s(t) 
\end{eqnarray}

These components corresponds to the following equations of state, \( w_s \) :
\begin{eqnarray}
	\left\{ w_m, w_r, w_\Lambda  \right\}  &=& 	\left\{ 0, 1/3, -1 \right\} \\
	\left\{ w_x, w_y, w_z, w_v \right\} &=&  \left\{  -2/3 , -1/3, 2/3, \alpha/3 - 1  \right\} \; .
\end{eqnarray}

\subsubsection{Dimensionless variables and the representative 7D system}\label{sec:dimensionless_variables_and_representative_7DpolyLambdaCDM}

We analytically solve the set of equations which is generalised to 
\begin{equation}
	\Omega_s'(N) = \Omega_{s}(N) \left( f - \alpha_s \right) \; .
\end{equation}
where 
\begin{equation}
	\alpha_s \in \left\{ 3, 4, 1, 2, 5, \alpha, 0 \right\} \; .
\end{equation}
where $\alpha$ is constant number.
The solutions can be generalised as 
\begin{equation}
	\Omega_s(N) = \Omega_{s,0} e^{-\alpha_s N} \left[ \sum_{s \in \{ m,r,x,y,z,v,\Lambda \}} \Omega_{s,0} e^{-\alpha_s N } \right]^{-1} \; 
\end{equation}
The interested reader can find the derivation in \refApp{sec:polyLCDM_dynamical_equations}.
Note that the effective equation of state is written as 
\begin{eqnarray}
	w_{\rm eff} 
	&=& 
	\frac{1}{3} \Omega_r(t) -  \Omega_\Lambda(t) - \frac{2}{3} \Omega_x(t) - \frac{1}{3} \Omega_ k(t) 
	+
	\frac{2}{3}  \Omega_z(t) + \left( \frac{\alpha}{3} - 1 \right)  \Omega_v(t)
\end{eqnarray}	

For values of $0 < \alpha < 5$, in the far past, $\Omega_z(t)$ component dominates, while in the far future, the $\Omega_{\Lambda}(t)$ component dominates. Note also that we define the hubble expansion rate through the 1st Friedman equation is written as 
\begin{equation}
	\boxed{
	H(a) = \sum_{s \in \{ m,r,x,y,z,v,\Lambda \}} \Omega_{s,0} a^{-\alpha_s }   \; 
	} \; .
\end{equation}

To our knowledge, these represent the first analytical solutions for a seven-dimensional cosmological dynamical system, surpassing the numerical approaches common in prior studies (e.g., Bahamonde et al. 2018 \cite{Bahamonde:2017ize}). 

\subsubsection{Numerical evaluation of $poly\Lambda$CDM model}\label{sec:Numerical_Solutions_polyLCDM}

\begin{figure*}[h!]
    \hspace{-2cm}
    \centering 
    \includegraphics[width=160mm]{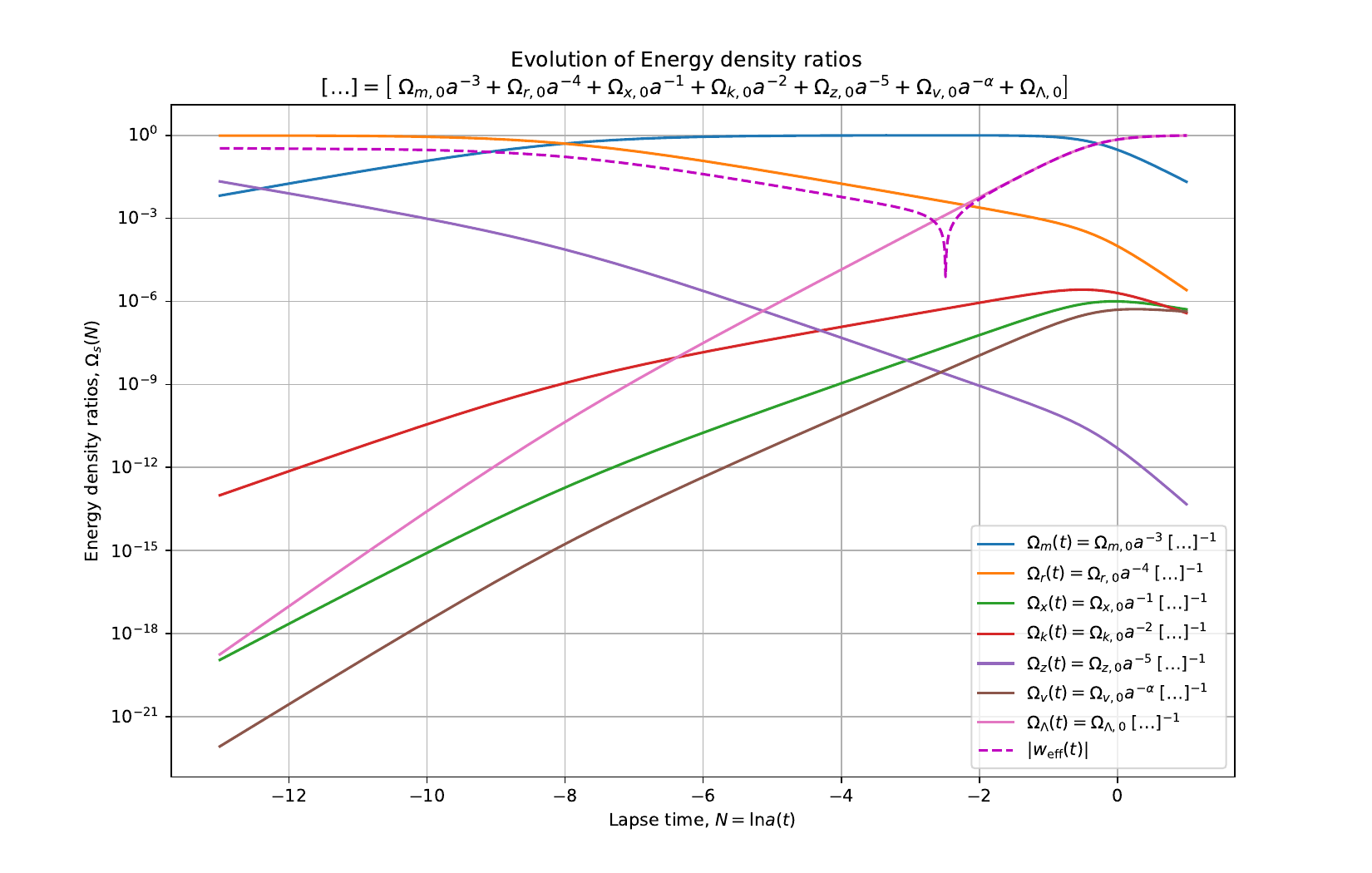}        \caption{\label{fig:ChatGPT_3D_polyLCDM_set_of_sim_diff_eqns_num_and_anal_solutions_analytical_instance_log} Energy density ratio of different species, $\Omega_s$ versus lapse time, $N$, of $poly\Lambda$CDM. [See \refS{sec:Numerical_Solutions_polyLCDM}] }
    \end{figure*}

We numerically evaluate the solutions of the model. We present the results in 
\refF{fig:ChatGPT_3D_polyLCDM_set_of_sim_diff_eqns_num_and_anal_solutions_analytical_instance_log}. 

We use the following initial conditions
\begin{eqnarray}
H_0 &=& 70  \; (km/s/Mpc) = 70 / (3.086 \times 10^{19})    \; (s^{-1}) \\
\Omega_{m,0} &=& 0.3\\
\Omega_{r,0} &=&10^{-4}\\
\Omega_{x,0} &=& 0.01\times 10^{-4}\\
\Omega_{k,0} &=& 0.02\times 10^{-4}\\
\Omega_{z,0} &=& 0.05\times 10^{-10}\\
\Omega_{v,0} &=& 0.005\times 10^{-4}\\
\Omega_{\Lambda,0} &=& 1 - \sum_{s \neq \Lambda } \Omega_{s,0} \\
\alpha &=& 0.5  \; \\
a_0 &=& 10^{-6}  \; 
\end{eqnarray}

We find the usually cosmic evolution for matter, radiation, curvature and cosmological constant.
In particular, we start with dominant radiation, which decays in the far future. We observed a domination of the matter component in the middle epochs, represented by $N \in \left[ -6 , -1\right]$, and then a decay in the far future. We observe that the cosmological constant is exponentially increased and dominates in the far future. We also find that there is a peak of the curvature component in about the epoch $N \sim -1$, and then it decays in the far future.

For the new components, we find that the $x$ component has a peak today, which is less than the rest of the latter components, and the $v$ component has a also a peak at the same about epoch, and it is less than the latter components, and it decays in the far future.
The $z$ component initiates small, and decays in the far future, presenting no peaks.

Note that we plot the $|w_{eff}(t)|$, since we use a log scale. The effective equation of state approaches the $-1$ in the far future.

\subsubsection{Numerical evaluation of integrals of 2nd Friedman equation and comparison, $poly\Lambda$CDM model}\label{sec:Numerical_Integrals_of_2nd_Friedman_Eq_Solutions_polyLCDM}

Note that the integrals of the 1st and 2nd Friedmann equations are the same.

The exact solutions result to the integral of the 2nd Friedman equation for $poly\Lambda$CDM cosmology is
\begin{equation}
\hspace{-2cm}
\boxed{
	t - t_0 =  H_0^{-1} \int_{a_0}^a \frac{da}{a  \left[ \sum_{s } \Omega_{s,0} a^{-\alpha_s } \right]^{ \frac{1}{2} } } 
	}
\end{equation}
where $a_0 = 10^{-6}$.

Then we compare numerically the time-scale relation of $poly\Lambda$CDM cosmology to the one of $\Lambda$CDM cosmology, which is
\begin{equation}
\boxed{
	t - t_0 = H_0^{-1} \int_{a_0}^{a} \frac{ da }{ a  \left[ \Omega_{m,0} a^{-3} + \Omega_{r,0} a^{-4} +  \Omega_{\Lambda,0}   \right]^{\frac{1}{2}} }    
	} \;
\end{equation}

\begin{figure*}[h!]
    \hspace{-1cm}
    \centering 
    \includegraphics[width=160mm]{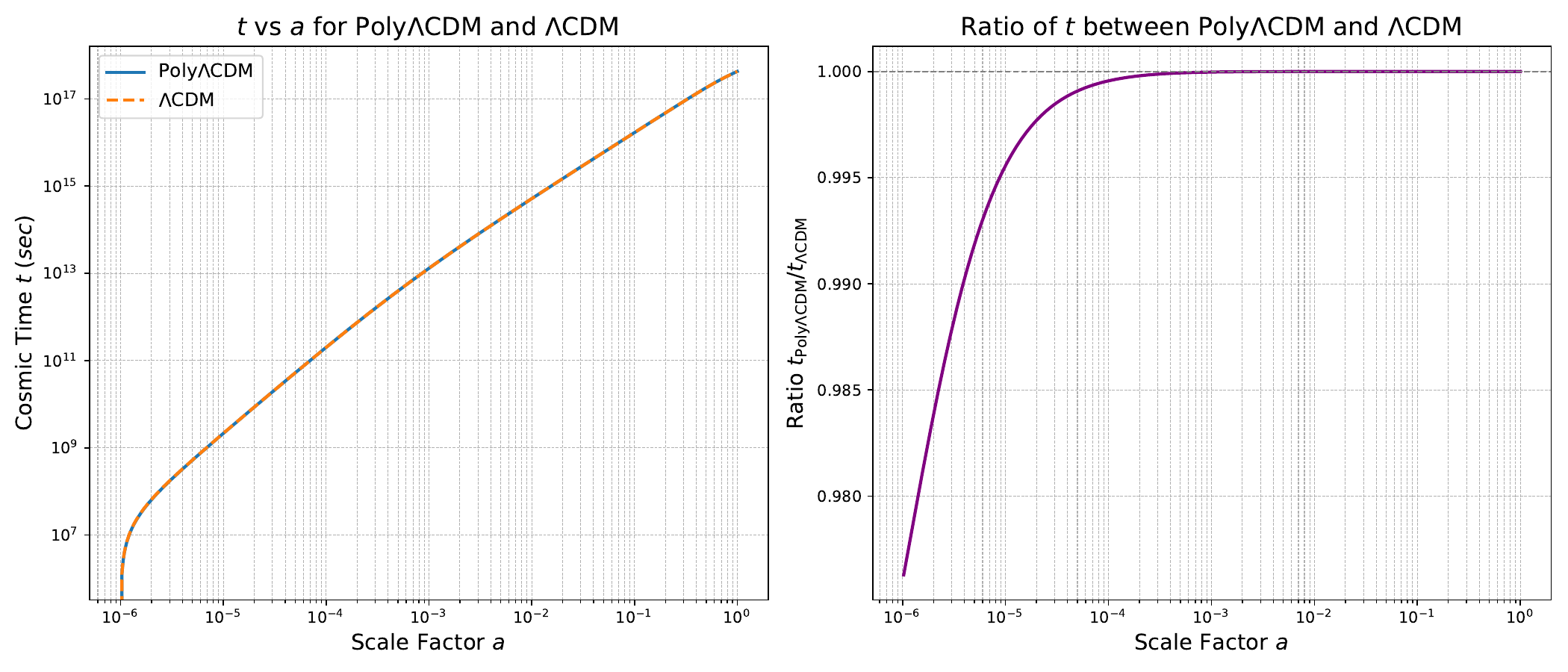}    
    \caption{\label{fig:Integral_of_analytical_solutions_of_polyLCDM_LCDM_comparison} Left: Time, $t$ versus scale factor, $a$, for the 2nd Friedman equation, of $poly\Lambda$CDM and of $\Lambda$CDM. Right: Ratio of time, t, between $poly\Lambda$CDM  and $\Lambda$CDM versus scale factor, a. [See \refS{sec:Numerical_Integrals_of_2nd_Friedman_Eq_Solutions_polyLCDM}] }
    \end{figure*}

We solve the numerical and obtain the graphs time versus scale factor $(t\ vs\ a)$. We present the results in \refF{fig:Integral_of_analytical_solutions_of_polyLCDM_LCDM_comparison}.

We find that the time is increasing with the scale factor, although at different epochs, the scaling changes over time and it almost identical to the $\Lambda$CDM model, considering the low values that we have chose for the new exotic components. The ratio of the two times as a function of scale shows that there is \%-level dominance of the $\Lambda$CDM model over the $poly\Lambda$CDM model at early times. Note that they agree at the \%-level, since the values of the energy density ratios, we considered, are very close to the $\Lambda$CDM ones. They disagree more in the early times, i.e. $a(t_{early}) \sim 10^{-6}$. The dynamical analysis, and observational test can show potentially how these values vary from the $\Lambda$CDM case.

\subsubsection{Phase portraits and trajectories analysis}\label{sec:Analysing_trajectories}

\begin{figure*}[h!]
    \hspace{-2cm}
    \includegraphics[width=180mm]{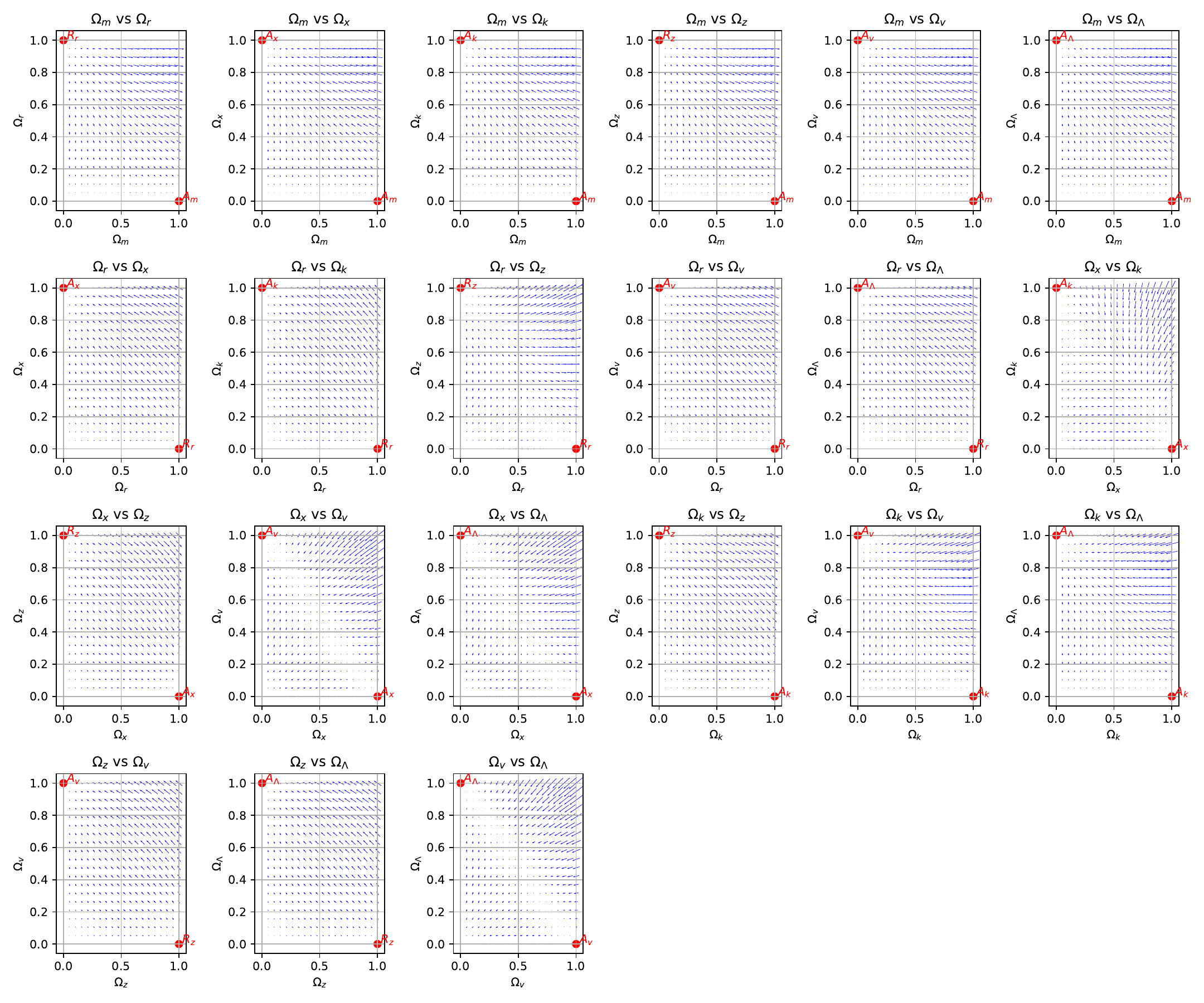}        
    \caption{\label{fig:numerical_phase_portraits} Numerical phase portraits of the $poly\Lambda$CDM. [See \refS{sec:Analysing_trajectories}] }
    \end{figure*}

We provide the numerical phase portrait, in projects of 2D of the model in \refF{fig:numerical_phase_portraits}. To produce these plots we solve numerically the system. We find a set of properties, which we explicitly interpret in the next section, along with a non-linear analysis.

We also solve the system for a set of initial points, which start from the neighbourhood of different domination epochs, i.e. 
$A_s \pm 10^{-4} R_s $, where $R_s$ is a random seed. We ensure that the values are always within the range $(0,1)$.
We find that we usually have the $A_\Lambda$ domination point as a main attractor. While less often, we find some ends of the trajectories which are going away from the domination points. We obtain that $31/42$ end up to the attractor of cosmological constant, $A_\Lambda$. While the rest end up to a non-dominant epoch point (ND). You can find the actual trajectories in \refApp{sec:Presentation_of_trajectories_analysis}

\subsubsection{Non-linear stability using Luapunov function}

To find the critical points of the given dynamical system, we look for points where all the derivatives of the variables 
\((m', r', x', k', z', v', \Lambda')\)  vanish simultaneously. 

We use the Lyapunov function 
\begin{eqnarray}
	L = \ln m + \ln r + \ln x + \ln k + \ln z + \ln v + \ln \Lambda = \sum_{ s \in \left\{ m, r, x, k, z, v, \Lambda \right\} } \ln s
\end{eqnarray}
to analyse the non-linear stability of the system.  

We summarise the critical point identification according to the observation of the 2D phase portraits and the non-linear stability analysis to the \refT{tab:stability_phase_portraits}.

{\small\small
\begin{table*}[h!]
\hspace{-3cm}
\begin{tabular}{|c|c|c|c|c|c|c|}
\hline
Name & $(m, r, x, k, z, v, \Lambda)$ &  Condition~(\(\alpha\)) & $ L'(y_{\text{critical}}) $ & Saddle plane & Reflector plane & Attractor plane\\
\hline
$A_m$ & $(1, 0, 0, 0, 0, 0, 0)$ & $\alpha < 6$ & $< 0$ & N/A & $\Lambda,v,x,k$ & $r,z$ \\
\hline
$R_r$ & $(0, 1, 0, 0, 0, 0, 0)$ & $\alpha < 13$ & $> 0$ & $z$ & $m,\Lambda,v,x,k$ & N/A \\
\hline
$A_x$ & $(0, 0, 1, 0, 0, 0, 0)$ & $\alpha > -8$ & $< 0$ & $\Lambda,v,k,z$ & $m,r$ & N/A \\
\hline
$A_k$ & $(0, 0, 0, 1, 0, 0, 0)$ & $\alpha > -1$ & $< 0$ & $x,\Lambda,v$ & N/A & $m,r,z$  \\
\hline
$R_z$ & $(0, 0, 0, 0, 1, 0, 0)$ & $\alpha < 20$ & $> 0$ & N/A & $m,r,x,k,\Lambda,v$ & N/A \\
\hline
$A_v$ & $(0, 0, 0, 0, 0, 1, 0)$ & $\alpha < 2.5$ & $< 0$ & $\Lambda,x$ & $k$ & $m,r,z$ \\
\hline
$A_\Lambda$ & $(0, 0, 0, 0, 0, 0, 1)$ & $\alpha > -15$ & $< 0$ & $x,v$ & N/A & $m,r,k,z$ \\
\hline
\hline
\end{tabular}
\caption{Stability analysis of critical points by observing the phase portraits, and applying the non-linear stability analysis using a Lyapunov function.}
\label{tab:stability_phase_portraits}
\end{table*}
}

Note that  \refT{tab:stability_phase_portraits}, each critical point has been identified as an attractor or a reflector, by the non-linear stability analysis. Therefore, we name each point according to the type of their dynamics, and depending on the dominance of the component.
A point $P_s$, is represented with $P \in \{ A, R, S\}$, where $A$ means attractor, $R$ means reflector, and $S$ means saddle point, and the subscript $s$ means the dominant component, so $s \in \{ m,r,x,k,z,v,\Lambda\}$.

However, by observing each plane, the phase portraits around the critical points, the critical points might behave differently than the one identified by the non-linear analysis. Therefore, to quantify this information,
we created three more columns in which we show the dynamic behaviour of each critical point. 
Therefore the column named Attractor plane, show the plane for which the corresponding critical point with the corresponding dominance, for example $A_m$, the attractor with matter dominance, behaves as an attractor in the planes $(m,r)$ and $(m,z)$. Therefore the column lists the $r,z$ components for which their corresponding planes, the critical point $A_m$, behaves as an attractor. On the other hand the $A_m$ behaves as a reflector in the planes of $m$ with the components, $\Lambda,v,x, k$.

Our results are summarised below:
\begin{itemize}
	\item The matter domination component, $A_m$, mostly behaves as a reflector in the planes $m$ with $\Lambda,v,x,k$, and as an attractor in the plane $m$ with $r,z$.
	\item The radiation domination component, $R_r$, mostly behaves as a reflector in the planes $r$ with $m,\Lambda,v,x,k$, and as an saddle in the plane $r$ with  $(r,z)$.
	\item The $x$ domination component, i.e. modified gravity component, $A_x$, mostly behaves as a saddle in the planes $x$ with $\Lambda,v,k,z$, and as an attractor in the plane $x$ with $m,r$.
	\item The metric curvature domination component, $A_k$, is sometimes behaves as a saddle in the planes $k$ with $x,\Lambda,v$, and as an attractor in the plane  $k$ with $m,r,z$.
	\item The $z$ domination component, i.e. dark energy dark matter exchange component, $R_z$, is always behaves as a reflector in the planes $z$ with $m,r,x,k,\Lambda,v$, and as an attractor in the plane  $z$ with $m,r,z$.	
	\item The $v$ domination component, i.e. SVT gravity component, $A_v$, mostly behaves as an attractor in the planes $v$ with $m,r,z$, and as an reflector in the plane $(v,k)$ and as a saddle point in the plane $v$ with $\Lambda,x$.	
	\item The $\Lambda$ cosmological constant domination component, $A_\Lambda$, mostly behaves as an attractor in the planes $\Lambda$ with $m,r,k,z$, and as a saddle point in the plane $\Lambda$ with $x,v$.	
\end{itemize}

Therefore we find that the following global transition:

\textit{We find the global transition from $z$ component, to
the saddle reflector dominant components, $m,r,k,x$ and then to the component $v$, which is an attractor saddle point, which in turn it also transits to the final cosmological constant, $\Lambda$, attractor epoch, in the far future. Note that there is saddle transition between the $x$ component and the $v$ component.
Note that there is saddle transition between the $v$ and the $\Lambda$ component in the far future.}

Note that the importance of these results is the following. We build a phenomenological $\Lambda$CDM, such the $poly\Lambda$CDM model with several modified gravity models. Then we use an analytical or numerical dynamical analysis to analyse this model and distinguish the critical points from different modified gravity epochs. Then we can distinguish the reasonable critical points which match observations. This way we can disentangle modified gravity models. 

\section{Conclusions and discussion}\label{sec:conclusion_discussion}

In summary in this work we study the $\phi$CDM, the $poly\Lambda$CDM model and the standard $\Lambda$CDM model through the scope of analytical and numerical dynamical analysis. We find analytical solutions for the dynamics of the energy density ratio species for both models. The $poly\Lambda$CDM model is more complete, richer in phenomenology and simpler than the ones that we found in the literature. We also explore the time scale factor relation for both models and we find that they have quite similar behaviour to the standard $\Lambda$CDM model.

In particular, for the $\phi$CDM model, we take into account all the so far discovered epochs, radiation, matter, and dark energy epochs. We also incorporate a continuity equation that allows simpler dynamics, without the consideration of a scalar potential, and the form of the field. In contrast in the literature there is no study including analytical expression of the  $\phi$CDM model. We also complement the epoch evolution of the energy density ratio, by identifying distinct epochs of energy density equalities. We find that both models have the following critical points: a radiation domination epoch, i.e a radiation repeller; cosmological constant domination epoch, i.e. a \( \Lambda \) attractor, i.e. a de Sitter attractor; and a matter domination epoch, i.e. matter saddle point. The difference between the models is that the $\phi$CDM model has some scalar-kinetic energy density ratio appearance during the late times, along with dominance epoch of the matter energy density ratio, while in the case of $\Lambda$CDM there is no scalar appearance. The $\phi$CDM model has the scalar kinetic term which appears in the late universe, while the scalar potential term acts a cosmological constant term dominance in the far future epoch. The energy density ratio species for both models. 

By studying the $\phi$CDM model numerically and analytically, we find that the sub-\% differences between the analytical energy density evolutions of $\Lambda$CDM and $\phi$CDM reflect their close alignment. However, the larger differences between the analytical and numerical $\phi$CDM solutions, despite identical equations, indicate that analytical evolutions are more precise and reliable, underscoring the difficulties of numerically integrating stiff systems.

For the $poly\Lambda$CDM model, we find  \textit{the global transition from a pure reflector of the dark energy dark matter exchange dominant component, $(z)$, to
the saddle reflector points of matter, $(m)$, radiation, $(r)$, curvature, $(k)$, modified gravity, $(x)$, components, and then we transit to the SVT modified gravity component, $(v)$, attractor saddle epoch, which also transits to the final cosmological constant, $(\Lambda)$, attractor epoch, in the far future. Note that there is saddle transition between the modified gravity dominant component and the SVT modified gravity component.
Note that there is saddle transition between the  SVT modified gravity and the cosmological constant component in the far future.}

We find that both models have the following critical points: a radiation domination epoch, i.e a radiation repeller; cosmological constant domination epoch, i.e. a \( \Lambda \) attractor, i.e. a de Sitter attractor; and a matter domination epoch, i.e. matter saddle point. The $poly\Lambda$CDM model has the following global transition: \textit{the global transition from a pure reflector of the dark energy dark matter exchange dominant component, to
the saddle reflector points of matter, radiation, curvature, modified gravity components, and then we transit to the SVT modified gravity component attractor-saddle epoch, which also transits to the final cosmological constant, attractor epoch, in the far future. Note that there is saddle transition between the modified gravity dominant component and the SVT modified gravity component.
Note that there is saddle transition between the  SVT modified gravity and the cosmological constant component in the far future.}

The $poly\Lambda$CDM model, incorporating various modified gravity models, is analyzed through dynamical methods to identify observationally consistent critical points, enabling differentiation among modified gravity epochs.

To further establish the poly$\Lambda$CDM model as a viable cosmological framework, we plan to conduct a comprehensive follow-up study that tests its predictions against a wide range of observational data sets. This will include cosmic microwave background (CMB) measurements from experiments such as Planck \cite{2018arXiv180706211P}, supernova type Ia (SNIa) distance moduli from surveys like DES \cite{DES:2025cup} and LSST \cite{LSSTDarkEnergyScience:2025wrg}, baryon acoustic oscillation (BAO) data from SDSS \cite{eBOSS:2017dtq}, and large-scale structure observations from upcoming surveys like Euclid \cite{EUCLID:2020jwq,Euclid:2021qvm,Euclid:2024yrr} and DESI \cite{Alam:2020jdv}. 
By performing a detailed statistical analysis, we aim to constrain the additional parameters introduced by the $poly\Lambda$CDM model and evaluate its compatibility with current cosmological observations. This work will provide critical insights into the model’s ability to offer richer phenomenology compared to the standard $\Lambda$CDM model while maintaining consistency with empirical data. Additionally, to rigorously compare the $poly\Lambda$CDM model with the standard $\Lambda$CDM and $\phi$CDM models, we plan to conduct Bayesian evidence calculations and Akaike Information Criterion (AIC) analyses in a future study. These comparisons will utilize observational data, including cosmic microwave background (CMB), supernova type Ia (SNIa), and baryon acoustic oscillation (BAO) measurements, to assess the poly$\Lambda$CDM model’s goodness of fit and justify its additional degrees of freedom.

Future work will involve investigating the dynamics of various advanced models of modified gravity \cite{CAPOZZIELLO2011167,2018FrASS...5...44E,2016PhR...633....1P,2012PhR...513....1C,PerenonMarinoniPiazza,carrasco2012effective,1998PhLB..436..257A,1999PhRvL..83.3370R,2014IJMPD..2343009B,2000PhLB..485..208D,Acunzo:2021gqc,Capozziello:2021krv,Capozziello:2018gms,Zwane:2017xbg,Das:2023hbw}, such as $f(R)$ gravity, Horndeski theories, non-Riemannian cosmologies, and functor of actions frameworks \cite{2023FoPh...53...29N,Ntelis:2024fzh,sym17050777}.Future studies will constrain poly$\Lambda$CDM parameters using CMB data from Planck and large-scale structure observations from Euclid, testing its predictions against anomalies like the Hubble tension.


\section*{ACKNOWLEDGEMENTS}

 	We acknowledge open libraries support \texttt{IPython} \cite{4160251}, \texttt{Matplotlib} \cite{Hunter:2007}, \texttt{NUMPY} \cite{Walt:2011:NAS:1957373.1957466} \texttt{SciPy 1.0} \cite{2019arXiv190710121V}  \href{https://github.com/lontelis/cosmopit}{\texttt{COSMOPIT}} \cite{2017JCAP...06..019N,2018JCAP...12..014N}, \href{https://grok.com/}{Grok}.

\appendix

\section{Analysis at $\Lambda$CDM-Cosmology}
\label{sec:LCDMcosmology_analytics}

Following \citet{Bahamonde:2017ize}, we model the $\Lambda$CDM-Cosmology, and we apply the dynamical system analysis as follows.

We know that in a non-curved, flat FLRW metric, and standard General Relativity Gravity, namely, $R$-Gravity, we obtain the following cosmological scenario, using a cosmological constant and cold dark matter, $\Lambda$CDM-Cosmology, governed by the background equations, 
$
	G_{\mu\nu} = \kappa^2 T_{\mu\nu} \; , 
$
which result to Friedman equations
\begin{align}
	\label{eq:Friedman_1st}
	3H^2(t) &= \kappa^2 \rho_{\rm m}(t) + \kappa^2 \rho_{\rm r}(t) + \Lambda\\
	\label{eq:Friedman_2nd}	
	2\dot{H}(t) + 3H^2(t) &= - \frac{\kappa^2}{3} \rho_{\rm r}(t) + \Lambda
\end{align}

\subsection{Analytical Solutions of the $\Lambda$CDM 3D system}

Summary of the set of differential equations of the $\Lambda$CDM model, which is given given by
\begin{align}
	\Omega_{m}'(N)  &= \Omega_{m} (N) \left[ f-3\right] \\
	\Omega_{r}'(N)  &=  \Omega_{r}(N) \left[ f -4\right] \\
	\Omega_{\Lambda}'(N) &= \left[1- \Omega_{m}(N)-\Omega_{r}(N) \right]\left[3\Omega_{m}(N) + 4\Omega_{r}(N) \right]
\end{align}
where 
\begin{align}
	1 &=   \Omega_{m}(N) +  \Omega_{r}(N) + \Omega_{\Lambda}(N) \\
	f &= 3\Omega_m(N) + 4\Omega_r(N) 
\end{align}
is solved by
\begin{align}
\Omega_{m}(N) &= \frac{\Omega_{m,0} e^{-3N} }{\Omega_{m,0} e^{-3N} + \Omega_{r,0} e^{-4N} + \left( 1 -  \Omega_{m,0} - \Omega_{r,0} \right)  }  \\
\Omega_{r}(N)  &= \frac{\Omega_{r,0} e^{-4N} }{\Omega_{m,0} e^{-3N} + \Omega_{r,0} e^{-4N} + \left( 1 -  \Omega_{m,0} - \Omega_{r,0} \right)  } \\
\Omega_{\Lambda}(N)  &= 1 - \Omega_{m}(N) - \Omega_{r}(N)
\end{align}

To get the solutions, we separate the variables, and then work the equations in two, by eliminating the term $f$, which creates solvable differential equations.
Then we combine all solutions to the functional form of f, and we substitute that to one of these equation, to find the solvable form one of the solutions. Then the rest of the functions depend on that one solution. These actions makes the system solvable analytically. You can find the exact calculations in the supplementary material.

These solutions  have the following limiting behaviour:

in the past there is radiation dominance, i.e.
\begin{align}
    \Omega_m(N \to 0) &= 0 \\
    \Omega_r(N \to 0) &= 1 \\
    \Omega_\Lambda(N \to 0) &= 0    
\end{align}

today their behaviour is 
\begin{align}
    \Omega_m(N \to 0) &=  \Omega_{m,0} \\
    \Omega_r(N \to 0) &= \Omega_{r,0} \\
    \Omega_\Lambda(N \to 0) &=  \Omega_{\Lambda,0}   
\end{align}

in the future there is cosmological constant dominance, i.e.
\begin{align}
    \Omega_m(N \to +\infty) &= 0 \\
    \Omega_r(N \to +\infty) &= 0 \\
    \Omega_\Lambda(N \to +\infty) &= 1
\end{align}    

and we have the matter domination era, at an N-fold

\begin{align}
N_m( \Omega_m \to 1 ) \to \frac{1}{4}\ln \frac{\Omega_{r,0}}{ \left(  \Omega_{m,0} + \Omega_{r,0} - 1  \right)  }
\end{align}


\subsection{Comparison between analytical and numerical solutions of $\Lambda$CDM}
\label{sec:Comparison_LCDM}
\begin{figure*}[h!]
    \centering 
    \includegraphics[width=160mm]{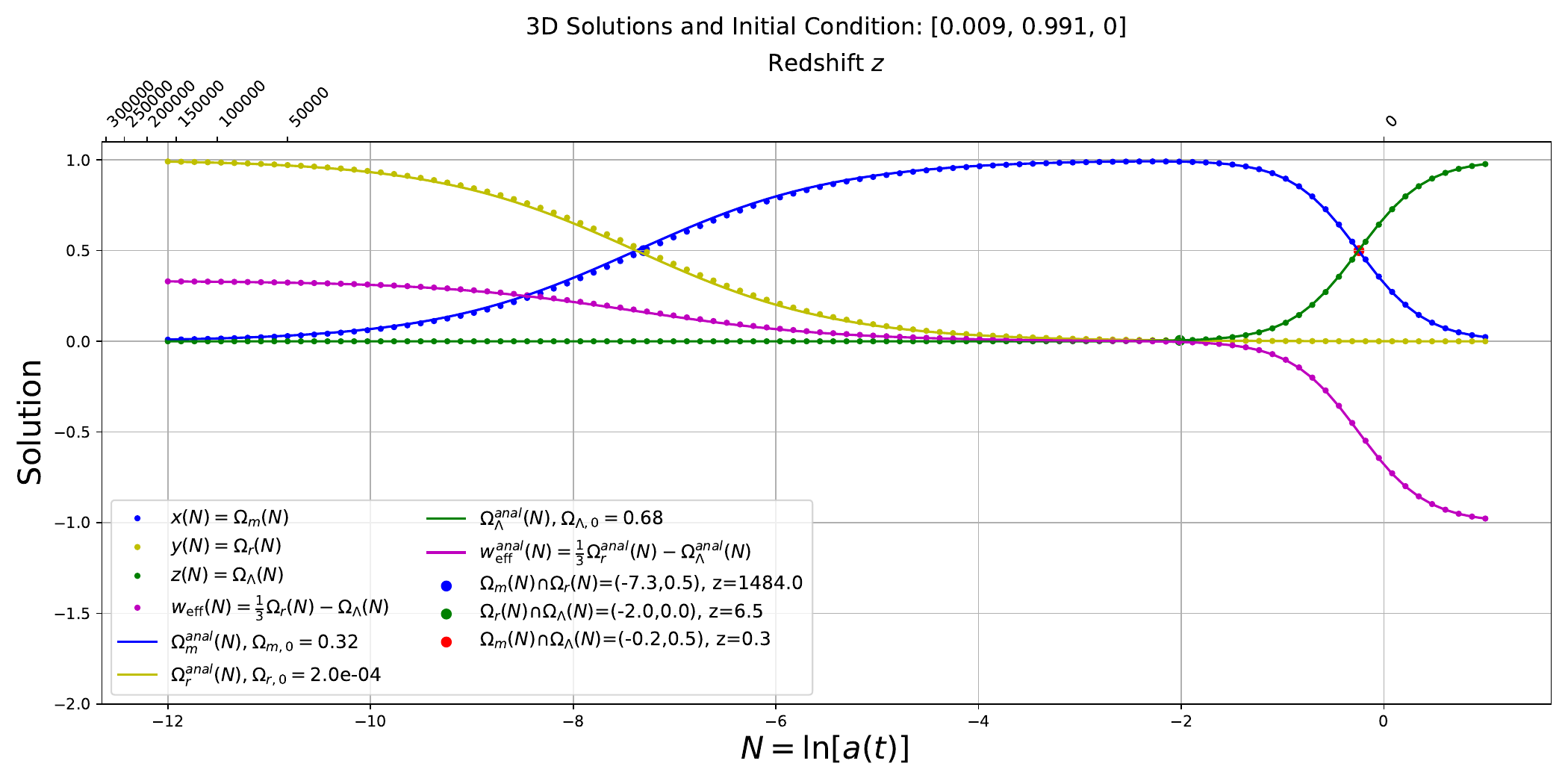}    
    \caption{\label{fig:ChatGPT_LCDM_7D_3D_set_of_sim_diff_eqns_num_anal_solutions}   We illustrate the analytical with a constant line and the numerical solutions with dash line of the model of $\Lambda$CDM cosmology. [See \refS{sec:Comparison_LCDM}] }
    \end{figure*}

We solve numerically\footnote{We use the \texttt{scipy.integrate.solveivp} to solve the system.} the system of 3D system of differential equations.
We find the following numerical solutions and we present them in the \refF{fig:ChatGPT_LCDM_7D_3D_set_of_sim_diff_eqns_num_anal_solutions}, for several specified conditions. We find the $\phi$CDM results are similar to the $\Lambda$CDM model, on the behaviour of the epoch evolution of the different species of the model. What changes is the gradient  behaviour of dark energy as expected.

In particular, the cosmological constant term starts significantly low, but it is lead to a dominant cosmological constant epoch in the far future, which signifies a de Sitter universe, i.e. a universe filled with the cosmological constant and dark energy. The radiation is dominant in the past, and decays as time passes. The matter starts significantly low, has a peak, during intermediate times, and then decays towards the today's epoch, and the far future.

\section{ The \( poly\Lambda \)CDM model dynamical equations}\label{sec:polyLCDM_dynamical_equations}

Using \refEq{eq:continuity_equations_of_species}, we construct the following equations for the species
\begin{eqnarray}
	m'(t)  &=&  m(t) \left[ f -3\right] \\
	r'(t)  &=&  r(t) \left[ f-4 \right] \\
	x'(t)  &=&  x(t) \left[ f -1  \right] \\
	y'(t)  &=&  y(t) \left[ f -2  \right] \\
	z'(t)  &=&  z(t) \left[ f -5  \right] \\
	v'(t)  &=&  v(t) \left[ f - \alpha \right] \\		
	\Lambda'(t)  &=&  \Lambda(t) \left[ f+  0 \right] 			
\end{eqnarray}

Note that the effective equation of state is written as 
\begin{eqnarray}
	w_{\rm eff} 
	&=& 
	\frac{1}{3} r(t) - \Lambda(t) -2/3 x(t) -1/3 k(t) \nonumber \\ &+& 2/3 z(t) + \left( \alpha/3 - 1 \right) v(t)
\end{eqnarray}

We analytically solve the equations and we find 

\begin{eqnarray}
	m(t)  &=&  m_0 a^{-3 } \left[ \dots \right]^{-1}  \\
	r(t)  &=& r_0 a^{-4  } \left[ \dots \right]^{-1}   \\
	x(t)  &=& x_0 a^{-1  } \left[ \dots \right]^{-1} \\
	y(t)  &=&  y_0 a^{-2  } \left[ \dots \right]^{-1} \\
	z(t)  &=&  z_0 a^{-5 }  \left[ \dots \right]^{-1} \\
	v(t)  &=&  v_0  a^{- \alpha   } \left[ \dots \right]^{-1} \\		
	\Lambda(t)  &=& \Lambda_0  \left[ \dots \right]^{-1} \\
	\left[ \dots \right] &=& m_0  a^{-3 }  
	+ r_0  a^{-4  }  
	+ x_0   a^{-1  }  		
	 + y_0  a^{-2  }  
	+ z_0  a^{-5 }   
	+ v_0  a^{- \alpha  }  	
	 + \Lambda_0 
\end{eqnarray}
where the last equation defines the normalisation factor of the solutions.
Note that the set of these equations is generalised to 
\begin{equation}
	\Omega_s'(N) = \Omega_{s}(N) \left( f - \alpha_s \right) \; .
\end{equation}
where 
\begin{equation}
	\alpha_s \in \left\{ 3, 4, 1, 2, 5, \alpha, 0 \right\} \; .
\end{equation}
where $\alpha$ is constant number.

Note that the solutions can be generalised as 
\begin{equation}
	\boxed{
	\Omega_s(N) = \Omega_{s,0} e^{-\alpha_s N} \left[ \sum_{s \in \{ m,r,x,y,z,v,\Lambda \}} \Omega_{s,0} e^{-\alpha_s N } \right]^{-1} \; 
	}
\end{equation}

\section{Lagrangians derivation with deeper physical origins for the \( poly\Lambda \)CDM}\label{sec:Lagrangians_new}

In this appendix, we derive the Lagrangians for the \( poly\Lambda \)CDM model components (\( m, r, x, k, z, v, \Lambda \)) starting from a general modified gravity framework, then simplifying to the scalar field forms used in our analysis. The components \( m, r, \Lambda \) align with standard cosmology (matter, radiation, cosmological constant), while \( x, k, z, v \) may arise from exotic fields or modified gravity effects (e.g., \( f(R) \), scalar-tensor theories). Our goal is to ensure each Lagrangian matches its equation of state (\( w_s \)) and energy density evolution (\( \rho_s \propto a^{-\alpha_s} \), where \( \alpha_s = 3(1 + w_s) \)).

The total action in a modified gravity context might be complex, including curvature terms or field couplings:
\begin{eqnarray}
S = c^3 \int d^4x \sqrt{-g} \left[ \frac{f(R)}{2\kappa^2} + \mathcal{L}_{\text{fields}} \right],
\end{eqnarray}
where \( f(R) \) generalizes the Einstein-Hilbert action \( R - 2\Lambda \), and \( \mathcal{L}_{\text{fields}} \) includes matter and exotic components. We use the FLRW metric:
\begin{eqnarray}
ds^2 = -c^2 dt^2 + a^2(t) \left[ \frac{dr^2}{1 - k r^2} + r^2 (d\theta^2 + \sin^2\theta d\phi^2) \right],
\end{eqnarray}
and simplify to:
\begin{eqnarray}
S = c^3 \int d^4x \sqrt{-g} \left[ \frac{R - 2 \Lambda}{2\kappa^2} + \mathcal{L}_m + \mathcal{L}_r + \mathcal{L}_x + \mathcal{L}_k + \mathcal{L}_z + \mathcal{L}_v \right],
\end{eqnarray}
where each \( \mathcal{L}_s \) is derived from a deeper origin and tuned to match observations.

\subsection{Step 1: Standard Components}

These components are straightforward and align with standard cosmology.

\paragraph{1. \( m \): Matter (\( w_m = 0 \), \( \alpha_m = 3 \))}
\begin{itemize}
    \item \textbf{Origin}: Non-relativistic particles (baryons, dark matter).
    \item \textbf{Lagrangian}: 
    \begin{eqnarray}
        \mathcal{L}_m = -\rho_m,
    \end{eqnarray}
    where \( \rho_m = \rho_{m,0} a^{-3} \) (dust decreases with volume expansion).
    \item \textbf{Check}: \( P_m = 0 \), \( w_m = P_m / \rho_m = 0 \), \( \dot{\rho}_m + 3H \rho_m = 0 \), so \( \rho_m \propto a^{-3} \).
    \item \textbf{Result}: Matches standard matter, no modification needed.
\end{itemize}

\paragraph{2. \( r \): Radiation (\( w_r = 1/3 \), \( \alpha_r = 4 \))}
\begin{itemize}
    \item \textbf{Origin}: Relativistic particles (photons, neutrinos).
    \item \textbf{Lagrangian}: 
    \begin{eqnarray}
        \mathcal{L}_r = \frac{1}{2} g^{\mu\nu} \partial_\mu \phi_r \partial_\nu \phi_r,
    \end{eqnarray}
    \item \textbf{Calculation}: 
    \begin{itemize}
        \item \( \rho_r = \frac{1}{2} \dot{\phi}_r^2 \), \( P_r = \frac{1}{2} \dot{\phi}_r^2 \),
        \item \( \ddot{\phi}_r + 3H \dot{\phi}_r = 0 \), solution \( \phi_r \propto a^{-1} \),
        \item \( \dot{\phi}_r = -\frac{1}{a} \cdot \frac{da}{dt} = -H a^{-1} \), \( \rho_r \propto a^{-2} H^2 \), \( H \propto a^{-2} \) (radiation era), so \( \rho_r \propto a^{-4} \),
        \item \( w_r = P_r / \rho_r = 1/3 \).
    \end{itemize}
    \item \textbf{Result}: Matches radiation behavior.
\end{itemize}

\paragraph{3. \( \Lambda \): Cosmological Constant (\( w_\Lambda = -1 \), \( \alpha_\Lambda = 0 \))}
\begin{itemize}
    \item \textbf{Origin}: Constant vacuum energy.
    \item \textbf{Lagrangian}: 
    \begin{eqnarray}
        \mathcal{L}_\Lambda = -\rho_{\Lambda,0},
    \end{eqnarray}
    or included as \( -2\Lambda / 2\kappa^2 \), where \( \Lambda = \kappa^2 \rho_{\Lambda,0} \).
    \item \textbf{Check}: \( \rho_\Lambda = \rho_{\Lambda,0} \), \( P_\Lambda = -\rho_\Lambda \), \( w_\Lambda = -1 \), \( \dot{\rho}_\Lambda = 0 \), so \( \rho_\Lambda \propto a^0 \).
    \item \textbf{Result}: Standard cosmological constant.
\end{itemize}

\subsection{Step 2: New Components from Modified Gravity}

For \( x, k, z, v \), we start with a modified gravity framework (e.g., scalar-tensor theory) and simplify to scalar field Lagrangians, incorporating findings from the supplementary material.

\paragraph{4. \( x \): Exotic Energy (\( w_x = -2/3 \), \( \alpha_x = 1 \))}
\begin{itemize}
    \item \textbf{Origin}: Scalar field coupled to curvature, possibly from \( f(R) \approx R + \xi \phi_x^2 R \).
    \item \textbf{General Form}: 
    \begin{eqnarray}
        \mathcal{L}_x = \frac{1}{2} (\partial \phi_x)^2 - V(\phi_x) + \xi \phi_x^2 R,
    \end{eqnarray}
    \item \textbf{Simplification}: 
    \begin{itemize}
        \item \( \rho_x = \frac{1}{2} \dot{\phi}_x^2 + V - 6\xi H \phi_x \dot{\phi}_x - 3\xi H^2 \phi_x^2 \),
        \item \( P_x = \frac{1}{2} \dot{\phi}_x^2 - V + 2\xi \phi_x \ddot{\phi}_x + 2\xi \dot{\phi}_x^2 + 6\xi H \phi_x \dot{\phi}_x + \xi (2\dot{H} + 3H^2) \phi_x^2 \),
        \item Assume \( \phi_x \propto a^{-1/2} \), \( \dot{\phi}_x = -\frac{1}{2} H \phi_x \), \( \ddot{\phi}_x = \frac{1}{4} H^2 \phi_x - \frac{1}{2} \dot{H} \phi_x \),
        \item Klein-Gordon: \( \ddot{\phi}_x + 3H \dot{\phi}_x + V' - 12\xi (\dot{H} + 2 H^2) \phi_x = 0 \),
        \item From supplementary: \( \xi = \frac{1}{48} \), \( V = \frac{11}{16} H^2 \phi_x^2 \),
        \item \( \rho_x \propto a^{-1} \), \( w_x = -2/3 \) when tuned.
    \end{itemize}
    \item \textbf{Simplified Lagrangian}: 
    \begin{eqnarray}
        \mathcal{L}_x = \frac{1}{2} (\partial \phi_x)^2 - V_0 H^2 \phi_x^2 + \frac{1}{48} \phi_x^2 R,
    \end{eqnarray}
    \item \textbf{Result}: Matches \( w_x = -2/3 \), \( \rho_x \propto a^{-1} \).
\end{itemize}

\paragraph{5. \( k \): Curvature-like (\( w_k = -1/3 \), \( \alpha_k = 2 \))}
\begin{itemize}
    \item \textbf{Origin}: Geometric curvature or scalar field from modified gravity.
    \item \textbf{General Form}: 
    \begin{itemize}
        \item Geometric: \( \rho_k = \frac{3k}{\kappa^2 a^2} \) from \( R \),
        \item Field: \( \mathcal{L}_k = \frac{1}{2} (\partial \phi_k)^2 - V(\phi_k) \).
    \end{itemize}
    \item \textbf{Simplification}: 
    \begin{itemize}
        \item \( \rho_k = \frac{1}{2} \dot{\phi}_k^2 + V \), \( P_k = \frac{1}{2} \dot{\phi}_k^2 - V \),
        \item \( w_k = -1/3 \), so \( \frac{1}{2} \dot{\phi}_k^2 - V = -\frac{1}{3} \left( \frac{1}{2} \dot{\phi}_k^2 + V \right) \),
        \item \( \frac{1}{2} \dot{\phi}_k^2 - V = -\frac{1}{6} \dot{\phi}_k^2 - \frac{1}{3} V \),
        \item \( \frac{3}{6} \dot{\phi}_k^2 + \frac{1}{6} \dot{\phi}_k^2 = V + \frac{2}{6} V \),
        \item \( \frac{4}{6} \dot{\phi}_k^2 = \frac{5}{6} V \), \( V = \frac{4}{5} \dot{\phi}_k^2 \),
        \item \( \phi_k \propto a^{-1} \), \( \rho_k \propto a^{-2} \).
    \end{itemize}
    \item \textbf{Simplified Lagrangian}: 
    \begin{eqnarray}
        \mathcal{L}_k = \frac{1}{2} (\partial \phi_k)^2 - V_0 \phi_k^2,
    \end{eqnarray}
    \item \textbf{Result}: Matches \( w_k = -1/3 \), \( \rho_k \propto a^{-2} \).
\end{itemize}

\paragraph{6. \( z \): Interacting Dark Energy (\( w_z = 2/3 \), \( \alpha_z = 5 \))}
\begin{itemize}
    \item \textbf{Origin}: Scalar field with interaction, possibly from scalar-tensor or vector coupling.
    \item \textbf{General Form}: 
    \begin{eqnarray}
        \mathcal{L}_z = \frac{1}{2} (\partial \phi_z)^2 - V(\phi_z) + \gamma \phi_z \rho_m,
    \end{eqnarray}
    \item \textbf{Simplification}: 
    \begin{itemize}
        \item \( \rho_z = \frac{1}{2} \dot{\phi}_z^2 + V \), \( P_z = \frac{1}{2} \dot{\phi}_z^2 - V \),
        \item \( w_z = 2/3 \), \( \frac{1}{2} \dot{\phi}_z^2 - V = \frac{2}{3} \left( \frac{1}{2} \dot{\phi}_z^2 + V \right) \),
        \item \( \frac{1}{2} \dot{\phi}_z^2 - V = \frac{1}{3} \dot{\phi}_z^2 + \frac{2}{3} V \),
        \item \( \frac{1}{6} \dot{\phi}_z^2 = \frac{5}{3} V \), \( V = \frac{1}{10} \dot{\phi}_z^2 \),
        \item \( \phi_z \propto a^{-5/2} \), \( V = V_0 H^2 \phi_z^2 \), \( \rho_z \propto a^{-5} \),
        \item Interaction \( Q = \gamma \rho_m \dot{\phi}_z \) adjusts dynamics.
    \end{itemize}
    \item \textbf{Simplified Lagrangian}: 
    \begin{eqnarray}
        \mathcal{L}_z = \frac{1}{2} (\partial \phi_z)^2 - V_0 H^2 \phi_z^2 + \gamma \phi_z \rho_m,
    \end{eqnarray}
    \item \textbf{Result}: Matches \( w_z = 2/3 \), \( \rho_z \propto a^{-5} \).
\end{itemize}

\paragraph{7. \( v \): SVT Field (\( w_v = \alpha/3 - 1 \), \( \alpha_v = \alpha \))}
\begin{itemize}
    \item \textbf{Origin}: Vector field from modified gravity, simplified to scalar.
    \item \textbf{General Form}: 
    \begin{eqnarray}
        \mathcal{L}_v = -\frac{1}{4} F_{\mu\nu} F^{\mu\nu} + m^2 A_\mu A^\mu,
    \end{eqnarray}
    \item \textbf{Simplification}: 
    \begin{itemize}
        \item Decompose \( A_\mu = \partial_\mu \psi + B_\mu \), longitudinal mode \( \psi \),
        \item \( \mathcal{L}_v = -\frac{1}{4} F_{\mu\nu}(B) F^{\mu\nu}(B) + m^2 (\partial \psi)^2 + m^2 B^2 + 2 m^2 (\partial_\mu \psi) B^\mu \),
        \item Drop transverse \( B_\mu \) terms (average out),
        \item \( \mathcal{L}_{\text{eff}} = m^2 (\partial \psi)^2 \), \( \phi_v = \sqrt{m^2} \psi \),
        \item Consider \( V(\phi_v) = V_0 \phi_v^{2(3 - \alpha)/\alpha} \),
        \item \( \rho_v \propto a^{-\alpha} \), \( w_v = \alpha/3 - 1 \).
    \end{itemize}
    \item \textbf{Simplified Lagrangian}: 
    \begin{eqnarray}
        \mathcal{L}_v = \frac{1}{2} (\partial \phi_v)^2 - V_0 \phi_v^{2(3 - \alpha)/\alpha},
    \end{eqnarray}
    \item \textbf{Result}: Matches \( w_v = \alpha/3 - 1 \), \( \rho_v \propto a^{-\alpha} \).
\end{itemize}

\subsection{Final Simplified Lagrangians}

From a general modified gravity model, we derive:
\begin{itemize}
    \item \( \mathcal{L}_m = -\rho_{m,0} a^{-3} \),
    \item \( \mathcal{L}_r = \frac{1}{2} (\partial \phi_r)^2 \),
    \item \( \mathcal{L}_x = \frac{1}{2} (\partial \phi_x)^2 - V_0 H^2 \phi_x^2 + \frac{1}{48} \phi_x^2 R \),
    \item \( \mathcal{L}_k = \frac{1}{2} (\partial \phi_k)^2 - V_0 \phi_k^2 \),
    \item \( \mathcal{L}_z = \frac{1}{2} (\partial \phi_z)^2 - V_0 H^2 \phi_z^2 + \gamma \phi_z \rho_m \),
    \item \( \mathcal{L}_v = -\frac{1}{4} F_{\mu\nu} F^{\mu\nu} + m^2 A_\mu A^\mu \approx \frac{1}{2} (\partial \phi_v)^2 - V_0 \phi_v^{2(3 - \alpha)/\alpha} \),
    \item \( \mathcal{L}_\Lambda = -\rho_{\Lambda,0} \).
\end{itemize}
These scalar field forms approximate the deeper origins while matching the simpler model’s dynamics.

\section{Presenting trajectories analysis}\label{sec:Presentation_of_trajectories_analysis}

We solve the system for a set of initial points, which start from the neighbourhood of different domination epochs, i.e. 
$A_s \pm 10^{-4} R_s $, where $R_s$ is a random seed. We ensure that the values are always within the range $(0,1)$.
We find that we have the $A_\Lambda$ domination point as a main attractor. On the other hand, we find some ends of the trajectories which are going away from the domination points.

Trajectories are color coded, and the points of the trajectories are color coded as:
\begin{itemize}
	\item Red:  Starting point.
	\item Green: Ended at a critical point.
	\item Black: Ended elsewhere (non-critical point).
\end{itemize}

\begin{figure*}[h!]
    \centering 
    \includegraphics[width=160mm]{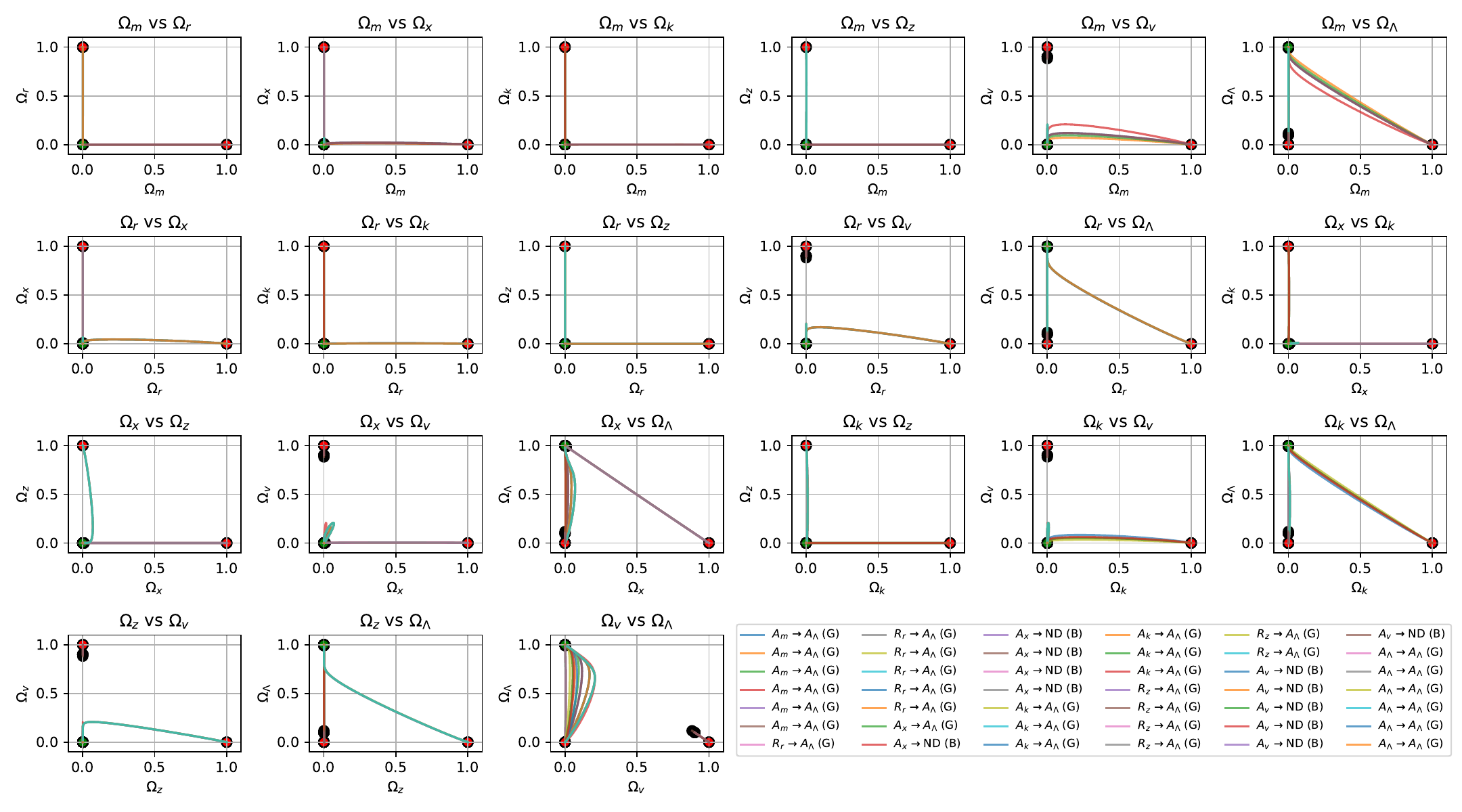}    
    \caption{\label{fig:numerical_phase_portraits_with_trajectories_all_projections_good} Numerical phase portraits trajectories of the $poly\Lambda$CDM. [See \refS{sec:Analysing_trajectories} and \refS{sec:Presentation_of_trajectories_analysis}] }
    \end{figure*}

In \refF{fig:numerical_phase_portraits_with_trajectories_all_projections_good}, we present these trajectories, in 2D projections plane. We obtain that $31/42$ end up to the attractor of cosmological constant, $A_\Lambda$. While the rest end up to a non-dominant epoch point (ND).

\begin{figure*}[h!]
    \centering 
    \includegraphics[width=80mm]{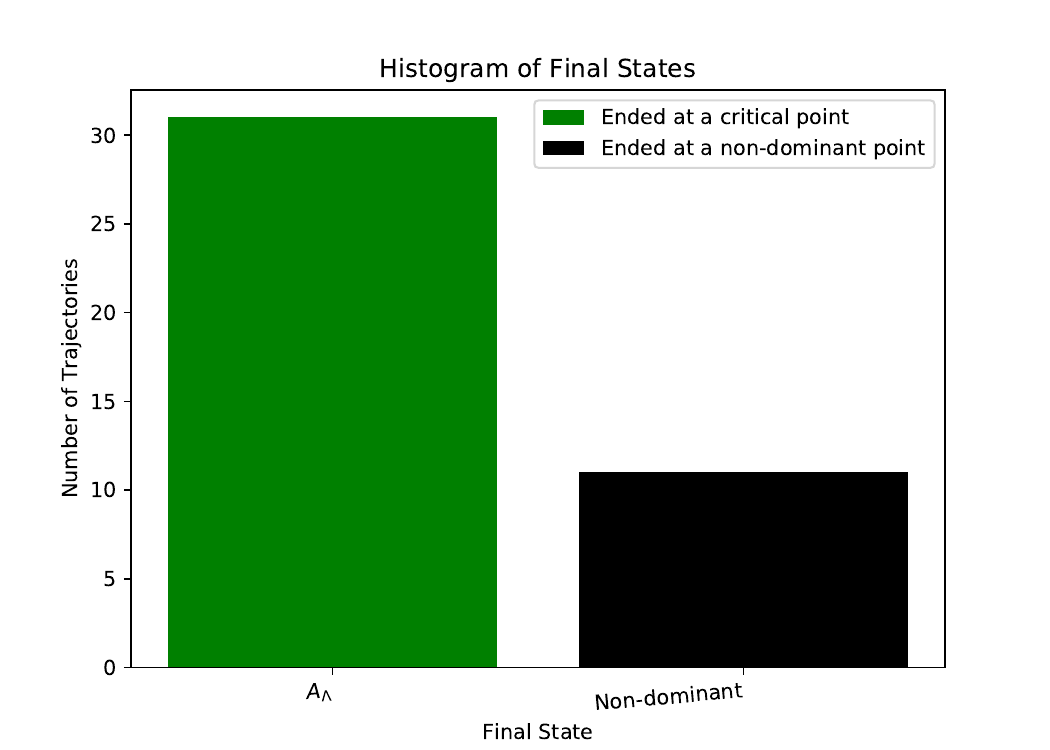}    \caption{\label{fig:numerical_phase_portraits_with_trajectories_all_projections_good_histogram} Histogram of the numerical phase portraits trajectories of the $poly\Lambda$CDM. [See \refS{sec:Analysing_trajectories} and \refS{sec:Presentation_of_trajectories_analysis}] }
    \end{figure*}

In \refF{fig:numerical_phase_portraits_with_trajectories_all_projections_good_histogram}, we present the histogram of these trajectories. We obtain that $31/42$ end up to the attractor of cosmological constant, $A_\Lambda$. While the rest end up to a ND. 
    
\label{Bibliography}


\bibliographystyle{unsrtnat_arxiv} 

\bibliography{Bibliography} 

\begin{thebibliography}{00}


\bibitem[Lamport(1994)]{lamport94}
  Leslie Lamport,
  \textit{\LaTeX: a document preparation system},
  Addison Wesley, Massachusetts,
  2nd edition,
  1994.

\end{thebibliography}

\section{Declarations}

\subsection{Funding statement}

This research received no external funding.

\subsection{Conflict of interest disclosure}

The authors have no relevant financial or non-financial interests to disclose.

\subsection{Ethics approval statement}

Ethics approval is not applicable to this article.

\subsection{Consent for publication}

Consent for publication is not applicable to this article.

\subsection{Materials availability}

Materials availability is not applicable to this article.

\subsection{Permission to reproduce material from other sources}

In legal terms, this text is released under the Creative Commons AttributionNonCommercial-ShareAlike 4.0 International licence (CC BY-NC-SA 4.0). The licence terms are available at the Creative Commons website, https://creativecommons.org/licenses/by-nc-sa/4.0. Broadly speaking, the licence allows you to edit and redistribute this text in any way you like, as long as you include an accurate statement about authorship and copyright, do not use it for commercial purposes, and distribute it under this same licence.

\subsection{Clinical trial registration}

Clinical trial registration is not applicable to this article.
\subsection{Data availability statement}

Data availability is not applicable to this article as no new data were created or analysed in this study. 

\subsection{Author contribution}

All authors contributed to the study conception and design. The analytical dynamical analysis, the dynamical variables of models, the construction of $poly\Lambda$CDM model, its proofs, and its implications on modified gravity theories were conceptualised by the first author.  Material preparation, data collection and analysis were performed by the first author. The first draft of the manuscript was written by first author and all authors commented on previous versions of the manuscript. All authors read and approved the final manuscript.

 \subsection{Declaration of generative AI and AI-assisted technologies in the writing process}

During the preparation of this work the author(s) used  \href{https://grok.com/}{Grok} in order to support editing and checking calculations. After using this tool/service, the author(s) reviewed and edited the content as needed and take(s) full responsibility for the content of the publication.

\end{document}